\documentstyle[12pt,epsfig]{article}                                                    

\newlength{\dinwidth}                                                    
\newlength{\dinmargin}                                                    
\def\lapproxeq{\lower .7ex\hbox{$\;\stackrel{\textstyle
<}{\sim}\;$}}
\def\gapproxeq{\lower .7ex\hbox{$\;\stackrel{\textstyle
>}{\sim}\;$}}
\def\gtrsim{\lower .7ex\hbox{$\;\stackrel{\textstyle
>}{\sim}\;$}}
\def\lesim{\lower .7ex\hbox{$\;\stackrel{\textstyle
<}{\sim}\;$}}
\def\be{\begin{equation}}
\def\ee{\end{equation}}
\def\bea{\begin{eqnarray}}
\def\eea{\end{eqnarray}}

\def\cc{c\bar{c}}
\def\qq{q\bar{q}}

\def\GeV{\rm GeV}
\def\a{{\alpha}_s}

\def\xb{{\bf x}}    
\def\yb{{\bf y}}    
\def\zb{{\bf z}}   

\begin{document}                                                    
\titlepage                                                    
\begin{flushright}                                                    
IPPP/08/03   \\
DCPT/08/06 \\                                                    
1 February 2008 \\                                                    
\end{flushright}                                                    
                                                    
\vspace*{2cm}                                                    
                                                    
\begin{center}                                                    
{\Large \bf Proton structure, Partons, QCD, DGLAP and beyond}                                                    
                                                    
\vspace*{1cm}                                                    
Alan D. Martin \\                                                    
                                                   
\vspace*{0.5cm}                                                    
Institute for Particle Physics Phenomenology, University of Durham, Durham, DH1 3LE            
\end{center}                                                    
                                                    
\vspace*{2cm}                                                    
                                                    
\begin{abstract}                                                    
We present an introductory discussion of deep-inelastic lepton-proton scattering as a means to probe the substructure of the proton. A r\'{e}sum\'{e} of QCD is given, emphasizing the running of the coupling constant and the DGLAP evolution equations for the parton densities. The determination of parton distributions is discussed and their importance for predictions of processes at the LHC is emphasized. Going beyond the pure DGLAP regime, we briefly discuss the behaviour of parton densities at low $x$, and the evidence for non-linear absorptive contributions.
\end{abstract}

\section{Deep inelastic scattering (DIS) introduced}
\begin{figure}
\begin{center}
\epsfxsize=20pc 
\epsfbox{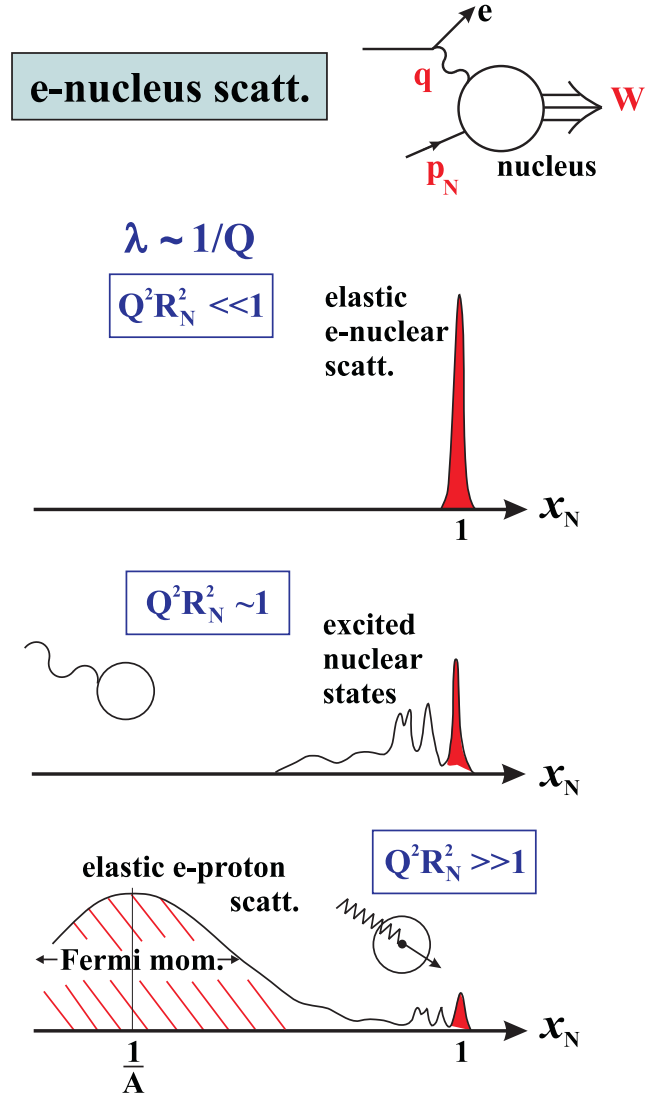} 
\caption{Electron-nucleus scattering, where $p_N$ and $q$ are the 4-momenta of the incoming nucleus and virtual photon respectively, and $W$ is the invariant mass of the outgoing hadronic system. The lower three diagrams are a schematic illustration of the cross section for electron-nucleus scattering, $eN \to eX$, plotted as a function of the scaling variable $x_N=Q^2/2p_N \cdot q$ at three different values of $Q^2$. In the lowest plot the wavelength $\lambda$ of the virtual photon probe is much less that the nuclear radius $R_N$, and the photon probes a constituent proton of the nucleus.  \label{fig:eN}}
\end{center}
\end{figure}

High energy electron scattering is an ideal probe of the structure of a composite object. For instance, consider the scattering of a beam of electrons on a nuclear target of mass $M_N$. The scattering occurs via the exchange of a virtual photon, see Fig.~\ref{fig:eN}. Since it is virtual, the photon is not on its mass shell. That is, its 4-momentum $q$ does not satisfy $q^2=0$. On the other hand, a real (ingoing or outgoing) particle or system must be on its mass shell. So the invariant mass $W$ of the outgoing system in Fig.~\ref{fig:eN} satisfies
\be
W^2~=~(p_N+q)^2~=~M_N^2+2p_N\cdot q+q^2,
\label{eq:W2}
\ee  
where $M_N$ and $p_N$ are the mass and 4-momentum of the nucleus. It follows that $q^2$ is negative. So we define $Q^2 \equiv -q^2$.

The wavelength of the probing photon $\lambda \sim 1/Q$. Let us follow what happens as we increase the electron energy, so that the photon probe has a shorter and shorter wavelength $\lambda$. We begin with $\lambda \gg R_N$, where $R_N$ is the radius of the nucleus. In this case the photon sees a ``point'' nucleus and we have {\it elastic} electron-nucleus scattering with $W=M_N$. Thus, from (\ref{eq:W2}),
\be
x_N~\equiv~\frac{Q^2}{2p_N\cdot q}~=~\left(\frac{Q^2}{2M_N\nu}\right)_{\rm lab.}~=~1,
\ee
where $\nu$ is the energy loss of the electron. The expression in the laboratory frame shows immediately that $Q^2 \equiv -q^2$ is positive. We sketch the corresponding elastic peak at $x_N=1$ in the first of the three plots of Fig.~\ref{fig:eN}. If we increase $Q$ until $\lambda \sim R_N$ then the outgoing system may be an excited nuclear state. Now $W>M_N$ and $x_N<1$, as shown in Fig.~\ref{fig:eN}. 

Finally, if $\lambda \ll R_N$, the photon may probe deep within the nucleus. The nucleus is broken up. We have deep $(Q^2 \gg M_N^2)$ inelastic $(W^2 \gg M_N^2)$ electron-nucleus scattering. Indeed, the electron may scatter off a constituent proton of the nucleus. In terms of $x_N$, the resulting electron-proton elastic scattering peak will occur at
\be
x_N~=~\frac{M}{M_N}\left(\frac{Q^2}{2M\nu}\right)_{\rm lab.}~=~\frac{1}{A},
\ee
but will be smeared out due to the Fermi momentum of the proton bound in the nucleus, see Fig.~\ref{fig:eN}. $M$ is the proton mass and $A$ is the number of nucleons in the nucleus.
The area under the Fermi-smeared peak gives the number of protons in the nucleus, and hence the position of the peak determines the number of neutrons. The reduction of the $eN$ elastic peak, with increasing $Q^2$, reflects the small chance of the $A-1$ spectator nucleons all happening to be moving in the direction of the outgoing struck proton and reforming the original nucleus. 

\begin{figure} [t]
\begin{center}
\epsfxsize=33pc 
\epsfbox{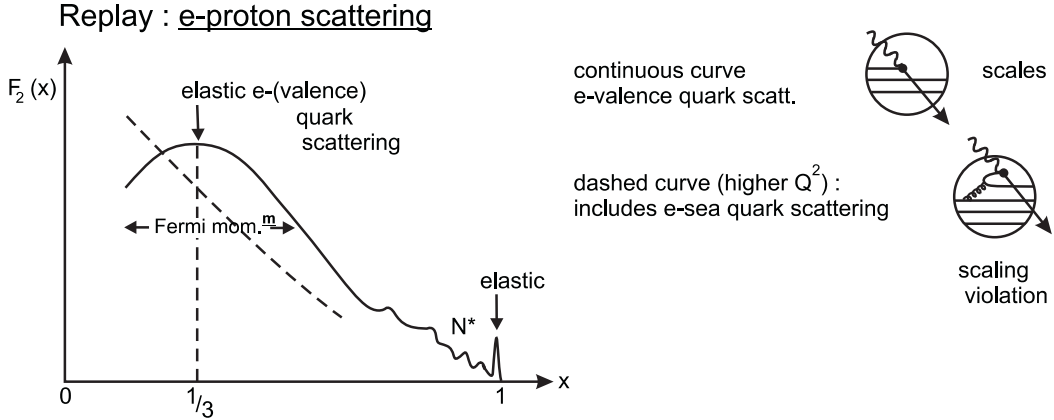} 
\caption{Schematic illustration of electron-proton scattering as a function of the Bjorken scaling variable $x\equiv Q^2/2p\cdot q$.  The proton structure function $F_2$ is defined in the next section.  The hadrons $N^*$ are excited states of the proton. If the proton consisted of just three valence point-like quarks the result would be the continuous curve independent of $Q^2$. However with increased resolution (higher $Q^2$) the photon may probe one of a pair of {\it sea} quarks produced from a radiated gluon via $g \to \qq$. Indeed, as $Q^2$ increases, the proton appears to have more and more constituents, which all must share its momentum, and so the distribution skews more and more towards small $x$. This trend from the continuous to the dashed curve is characteristic of QCD scaling violations. \label{fig:ep}}
\end{center}
\end{figure}
Let us increase $Q^2$ even further. Suppose that protons are made up of three point-like quarks, then high-energy electron-proton scattering will simply be a replay of electron-nucleus scattering one layer of substructure down. We have an analogous sequence of diagrams to those shown in Fig.~\ref{fig:eN}, but with $R_N$ replaced by the proton radius $R$. Also the scattering probabilities should now be plotted in terms of
\be
x~=~\left(\frac{Q^2}{2p\cdot q}\right),
\ee
where $p$ is the 4-momentum of the proton. The continuous curve in Fig.~\ref{fig:ep} is the analogue of the lowest plot in Fig.~\ref{fig:eN}. It shows the elastic $eq$-scattering peak Fermi-smeared about $x=1/3$, together with traces of the elastic $ep$ peak at $x\sim 1$. If there were no further substructure, this curve would persist as $Q^2$ increases. We would have (Bjorken) scaling; the scattering depends only on the ratio $x=Q^2/2p\cdot q$, and not on the two variables, $Q^2$ and $p\cdot q$, individually. $x$ is known as the Bjorken scaling variable.  

In summary, as $Q^2$ increases, we first have `nuclear' scaling with a peak at $x_N=1$, then violations of scaling, following by `proton' scaling with a peak at $x \sim 1$, followed by violations, and then `quark' scaling with a
peak at $x \sim 1/3$. If the quarks themselves had substructure then, as $Q^2$ increases even further, we would enter yet again a region of scaling violations followed by another onset of scaling.  But history does not seem to repeat itself. Scaling violations are observed but these reflect the field theory of quarks and gluons (QCD) with coupling $\a$.  The photon ``sees'' the proton made up of the three quarks (called {\it valence} quarks) and an arbitrary number of $\qq$ pairs (made up of {\it sea} quarks). The sea quarks originate from gluons, via $g \to \qq$, themselves radiated from quarks, see the sketch on the right of Fig.~\ref{fig:ep}.  Suppose the photon probes a quark carrying a fraction $\xi$ of the proton's momentum $p$. Then for essentially massless quarks we have
\be
(\xi p+q)^2~=~m^2_q~\simeq~0,~~~~~~{\rm that ~is}~~~~~\xi~\simeq~Q^2/2p\cdot q~=~x.
\ee
Consequently as $Q^2$ increases, more and more partons (that is quarks and gluons) become evident which have to share the momentum of the parent proton.  Each carries a smaller fraction $\xi =x$ of the momentum, and we get QCD scaling violations (which, as we will see, have the form $\a P {\rm log}(Q^2/\mu^2)$) as indicated by the dashed line in Fig.~\ref{fig:ep}. On hearing this for the first time from Wilczek, one of the discoverers of QCD, a famous experimentalist said  ``You expect us to measure logarithms? Not in your lifetime, young man''.  Yet today the high precision DIS data from HERA and earlier fixed-target experiments show exactly the QCD logarithmic scaling violations predicted. A collection of plots (which show the scaling violations) compiled from these deep inelastic $ep$ scattering data can be found in Section 16 of the Review of Particle Properties\cite{RPP}. Introductory discussions of DIS can be found, for example, in Refs.\cite{close,hm,rdf,roberts,altarelli,esw,dcs}.

\begin{figure}
\begin{center}
\epsfxsize=20pc 
\epsfbox{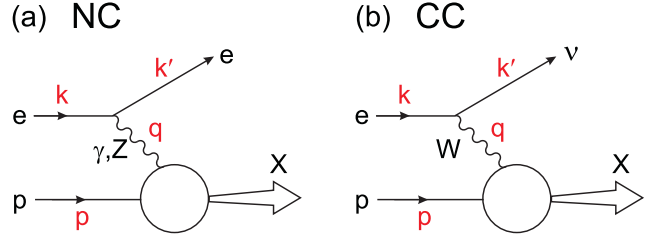} 
\caption{Neutral- and charged-current DIS mediated by ($\gamma ,Z$) and $W$ exchange respectively. \label{fig:ZW}}
\end{center}
\end{figure}
\section{The DIS observables: the structure functions}
The DIS process, $ep \to eX$, is shown in Fig.~\ref{fig:ZW}(a).  We talk of the neutral current (NC) DIS mediated by $\gamma$ and $Z$ exchange. We also have charged-current (CC) DIS mediated by $W$ exchange, shown in the second diagram. Recall that by ``deep'' we mean $Q^2 \gg M^2$ and by ``inelastic'' we mean $W^2=(p+q)^2 \gg M^2$.  

The NC cross section is of the form
\be
\frac{d\sigma}{dxdy}~=~xs\frac{d\sigma}{dxdQ^2}~=~\frac{2\pi y \alpha^2}{Q^4}~\sum_j \eta_j L^{\mu\nu}_j~W^j_{\mu\nu},
\label{eq:LW}
\ee
where the sum is over $j=\gamma,~Z$ and $\gamma Z$ representing photon and $Z$-boson exchange and the interference between them; and where
\be
\eta_\gamma=1,~~~~~~~~\eta_{\gamma Z}=\left(\frac{G_F M^2_Z}{2\sqrt{2}\pi \alpha}\right)\left(\frac{Q^2}{Q^2+M_Z^2}\right),~~~~~~~~~\eta_Z~=~\eta_{\gamma Z}^2.
\ee
We see the effects of the $\gamma$ and $Z$ propagators, and of the QED coupling $\alpha$ and the Fermi coupling $G_F$. Besides $x$ and $Q^2$, associated with the hadronic vertex, we have a variable ($y$ or $s$) which depends the energy of the whole $ep$ system
\be
y~=~\frac{p \cdot q}{p \cdot k}~=~\left(\frac{\nu}{E}\right)_{\rm lab.frame},~~~~~~~~~~~s~=~(k+p)^2~\simeq~\frac{Q^2}{xy}.
\ee
Both $x$ and $y$ must lie in the range from 0 to 1. The physical interpretation of $y$ is given in (\ref{eq:y}) below.

$L^{\mu\nu}$ is the tensor from the leptonic vertex {\it known} in terms of $k$ and $k'$, and $W_{\mu\nu}$ is the {\it unknown} tensor describing the hadronic vertex.  Although $W_{\mu\nu}$ is unknown it must be constructed from the 4-momenta $p,~q$ and the metric tensor $g_{\mu\nu}$. For unpolarised DIS, there are three tensor forms satisfying the requirements of current conservation $q^\mu W_{\mu\nu}=q^\nu W_{\mu\nu}=0$. In this case the general form is
\be
W_{\mu\nu}~=~\left(-g_{\mu\nu}+\frac{q_\mu q_\nu}{q^2}\right)F_1(x,Q^2)+\frac{\hat{P}_\mu \hat{P}_\nu}{p \cdot q}F_2(x,Q^2)-i\epsilon_{\mu\nu\alpha\beta} \frac{q^\alpha p^\beta}{2p \cdot q}F_3(x,Q^2),
\label{eq:W}
\ee
where $\hat{P}_\mu=p_\mu-(p \cdot q)q_\mu/q^2$.  The observable structure functions, $F_i(x,Q^2)$, are functions of two scalar variables $x$ and $Q^2$ which can be constructed from $p$ and $q$. Note that the last term, with a $\vec{q} \times \vec{p}$ type structure, does not conserve parity. Thus $F_3=0$ if $Z$ exchange is negligible. If we insert the general form (\ref{eq:W}) into (\ref{eq:LW}) and use the known forms of $L^{\mu\nu}$, then, after some algebraic manipulation, we find
\be
\frac{d\sigma}{dxdQ^2}~=~\frac{2\pi \alpha^2}{xQ^4}~(Y_+F_2\pm Y_-xF_3-y^2F_L)
\label{eq:X}
\ee
in the $M^2/Q^2 \to 0$ limit, where
\be
Y_{\pm}~=~1 \pm (1-y)^2~~~~~~~~{\rm and}~~~~~~~~~~~~~~F_L=F_2-2xF_1.
\ee 
A similar expression holds for CC DIS (that is $eN \to \nu X$ or $\nu N \to eX$). For both NC and CC processes, the $-$ sign for $Y_-$ is taken for an incoming $e^+$ or $\overline{\nu}$, and the $+$ sign is taken for an incoming $e^-$ or $\nu$.  Complete expressions for the lepton and hadron tensors $L_{\mu \nu},~W_{\mu \nu}$, the structure functions and the cross sections, including those for polarised DIS, can be found in Section 16 of the Review of Particle Properties\cite{RPP}.

For the moment let us focus on pure $\gamma$ exchange, so $F_3=0$. Even then to determine both $F_2$ and $F_L$ as functions of $x$ and $Q^2$ we need to measure the $y$ dependence. That is we need to perform DIS experiments at a range
of $ep$ energies\footnote{Indeed, the final data runs of HERA were made at lower energies specifically to enable direct measurements of $F_L$ to be performed. These should be available for DIS2008.}. We will see that $F_L=F_2-2xF_1\simeq 0$, so often the experiments either used QCD to calculate $F_L$ or assumed that it was zero, and presented results for $F_2(x,Q^2)$.  Nowadays, particularly with the advent of high $y$ data, the data are presented in terms of the so-called reduced cross section
\be
\sigma_{\rm red}(x,Q^2)~=~F_2(x,Q^2)-(y^2/Y_+)F_L(x,Q^2).
\ee

\begin{figure}[t]
\begin{center}
\epsfxsize=25pc 
\epsfbox{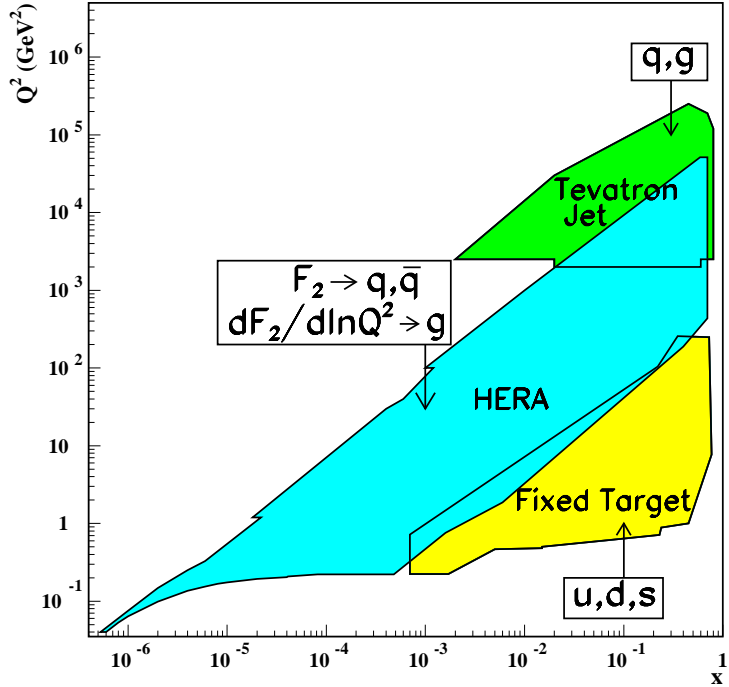} 
\caption{Kinematic domains in $x$ and $Q^2$ probed by fixed-target and collider experiments, shown together with the important constraints they make on the various parton distributions. \label{fig:kin}}
\end{center}
\end{figure}

The HERA machine at DESY collided 30 GeV electrons head-on with 920 GeV protons, giving
\be
s \simeq 4 E_e E_p \sim 10^5 ~{\rm GeV}^2. 
\ee
Thus $Q^2 \simeq xys \lapproxeq 10^5 ~{\rm GeV}^2$, and $x \simeq Q^2/ys \gapproxeq 10^{-4}$ for $Q^2=10~{\rm GeV}^2$. We see from Fig.~\ref{fig:kin} that the $(x, Q^2)$ reach of HERA is about two orders of magnitude better than the earlier fixed-target DIS data.
So given the data how do we interpret it?  Let us start with the Quark Parton Model.

\begin{figure}[t]
\begin{center}
\epsfxsize=25pc 
\epsfbox{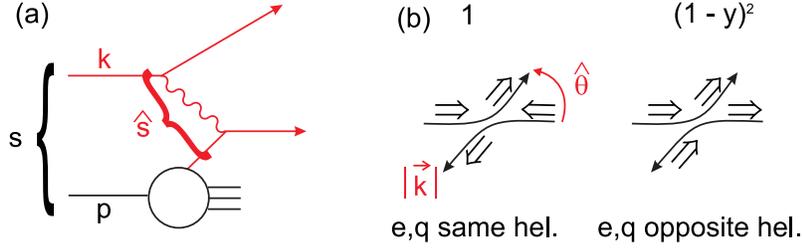} 
\caption{(a) DIS via the Quark Parton Model; the subprocess $eq \to eq$ occurs at c.m. energy $\sqrt{\hat s}$. (b) $eq$ scattering with equal (opposite) helicities occurs with weighting 1 ($(1-y)^2$) due to angular momentum effects. We also show the centre-of-mass scattering angle $\hat{\theta}$ and note that for essentially massless fermions all the particles have 3-momenta of magnitude $|\vec{k}|$. \label{fig:hel}}
\end{center}
\end{figure}

\section{The Quark Parton Model}
The basic idea of the QPM is that in the DIS process, $ep \to eX$, the virtual proton interacts with one of the quark constituents of the proton, see Fig.~\ref{fig:hel}(a). We view the process from a frame in which the proton is moving very fast so that the relativistic time dilation slows down the rate with which the quarks interact with each other. Thus the struck quark appears essentially free during the short time (about $1/Q$) that it interacts with the photon. As a result the $ep$ interaction may be written as an incoherent sum (of {\it probabilities}) of scattering from single {\it free} quarks
\be 
\frac{d\sigma}{dxdQ^2}~=~\sum_q \int^1_0 d\xi ~f_q(\xi)\left(\frac{d\hat{\sigma}_{eq}}{dxdQ^2}\right)~,
\label{eq:epqpm}
\ee
where $f_q(\xi)$ is the probability of finding the quark $q$ in the proton carrying a fraction $\xi$ of its momentum. The electron-quark cross section has the form
\be
\frac{d\hat{\sigma}_{eq}}{dxdQ^2}~=~\frac{2\pi \alpha^2 e^2_q}{\hat{s}^2} \left(\frac{\hat{s}^2 + \hat{u}^2}{\hat{t}^2}\right)\delta(x-\xi),
\label{eq:eqqpm}
\ee
where $\hat{s},~\hat{t}$ and $\hat{u}$ are the Mandelstam variables for the $eq \to eq$ subprocess.  Two sets of alternative expressions are
\be
\begin{array}{ll}
~~~\hat{s}=(xp+k)^2 \simeq 2x p \cdot k \simeq xs,~~~~~~  & \hat{s} \simeq 4\vec{k}^{~2}, \\
~~~\hat{t}=-Q^2\simeq -xys  & \hat{t} =-2\vec{k}^{~2} ~(1-{\rm cos}~\hat{\theta}) \\
~~~\hat{u} \simeq -\hat{s} -\hat{t} \simeq -x(1-y)s & \hat{u} =-2\vec{k}^{~2} ~(1+{\rm cos}~\hat{\theta}),
\end{array}
\ee
 where $|\vec{k}|$ and $\hat{\theta}$ are the magnitude of the $e,q$ three-momenta and the scattering angle in the $eq$ centre-of-mass frame. If we insert the first set into (\ref{eq:eqqpm}), then (\ref{eq:epqpm}) becomes
\be 
\frac{d\sigma}{dxdQ^2}~=~\frac{2\pi \alpha^2}{Q^4}\sum_q \int^1_0 d\xi ~f_q(\xi)~e^2_q ~\left[ 1+(1-y)^2\right]~\delta(x-\xi).
\label{eq:qpm}
\ee
Insight into the $y$ dependence is obtained by comparing the two sets of equations for $\hat{s},~\hat{t}$ and $\hat{u}$.
We see that
\be
y~=~\frac{1}{2}(1-{\rm cos}~\hat{\theta}),
\label{eq:y}
\ee
so $y=0$ corresponds to forward scattering and $y=1$ to backward scattering. If $e$ and $q$ have opposite helicities then there can be no backward $(\hat {\theta}=\pi)$ scattering by the conservation of $J_z$. This is the origin of the weighting $(1-y)^2$ in Fig.~\ref{fig:hel}(b). Crucial to this argument is the fact that at high energies $(E \gg m_{\rm fermion})$ the fermion helicity is conserved at a gauge boson vertex. 

If we re-write the QPM formula (\ref{eq:qpm}) in the form 
\be 
\frac{d\sigma}{dxdQ^2}~=~\frac{2\pi \alpha^2}{xQ^4}\sum_q \int^1_0 d\xi ~f_q(\xi)~e^2_q ~xY_+~\delta(x-\xi),
\label{eq:qpm1}
\ee
and then compare with the general structure function formula (\ref{eq:X}), assuming only $\gamma$-exchange, we obtain
\be
F_2~=~2xF_1~=~\sum_q\int_0^1 d\xi~f_q(\xi)~xe^2_q~ \delta(x-\xi) ~=~ \sum_q~e^2_q~xf_q(x).
\label{eq:QPMf}
\ee
The first equality (i.e. $F_L=0$) is known as the Callan-Gross relation, and holds because the quarks have spin 1/2. If the quarks had spin 0, then $F_1$ would have been 0. Also notice that in the QPM the structure functions scale, that is have no $Q^2$ dependence.

We have noted that the proton is made of valence quarks ($uud$) and sea quarks in $\qq$-pairs. When probed at a scale $Q$ all quark flavours with $m_q \lapproxeq Q$ are active. Usually the flavour is used as a shortened notation for a parton distribution. So, for example,
\be
\begin{array}{c}
f_u (x)~\equiv~u(x)~=~u_{\rm v}(x)+u_{\rm sea}(x), \\
f_{\bar {u}}(x)~\equiv~\bar{u}(x)~=~u_{\rm sea}(x).
\end{array}
\ee
We therefore have flavour sum rules
\be
\int^1_0(u-\bar{u})dx=\int_0^1 u_{\rm v}dx=2;~~~~~~~~ \\
\int^1_0(d-\bar{d})dx=\int_0^1 d_{\rm v}dx=1.
\ee

The structure function measurements can be used to reveal the quark flavour composition of the proton. From (\ref{eq:QPMf}) we have
\be 
F^{ep}_2~=~x\left(\frac{4}{9}u+\frac{1}{9}d+\frac{1}{9}s+...+\frac{4}{9}\bar{u}+\frac{1}{9}\bar{d}+\frac{1}{9}\bar{s}+... \right).
\label{eq:p}
\ee
Using isospin invariance it follows that the neutron structure function is
\be 
F^{en}_2~=~x\left(\frac{4}{9}d+\frac{1}{9}u+\frac{1}{9}s+...+\frac{4}{9}\bar{d}+\frac{1}{9}\bar{u}+\frac{1}{9}\bar{s}+... \right).
\label{eq:n}
\ee
Similar formulae can be obtained for CC DIS. For the $ep \to \nu X$ processes we have
\be
\frac{d\sigma (e^{\pm})}{dxdQ^2}~=~\frac{G_F^2}{2\pi x} \left(\frac{M_W^2}{Q^2+M_W^2}\right)^2~(Y_+F_2^W\mp Y_-xF_3^W-y^2F_L^W).
\label{eq:WCC}
\ee
Let us consider $e^-p \to \nu X$ (or $\bar{\nu} p \to e^+ X$); here the basic subprocesses are 
\be
\begin{array}{c}
e^- u \to \nu d, ~~~~~~~~{\rm with~ ``same"~ helicities} \\
e^-\bar{d} \to \nu \bar{u}, ~~~~~~~~{\rm with~``opposite" ~helicities.}
\end{array}
\ee
  If we now recall that $Y_{\pm} \equiv 1\pm (1-y)^2$, and use the helicity diagrams of Fig.~\ref{fig:hel}(b), then it follows that
\be
F_2^{W^-}=2x(u+\bar{d}+c+\bar{s}...),~~~~~~~~~~xF_3^{W^-}=2x(u-\bar{d}+c-\bar{s}...).
\label{eq:CCW}
\ee
For $e^+ p \to \bar{\nu} X$ (and $\nu p \to e^- X$) the structure functions are obtained by the flavour interchanges $d \leftrightarrow u$, $s \leftrightarrow c$, while those for the neutron are obtained from those of the proton by the interchange $u \leftrightarrow d$. Thus at large $x$, where the valence quark distribution dominates, we have
\be
\sigma^{\rm CC}(e^-p)~ \sim~ xu_{\rm v},~~~~~~~{\rm and}~~~~~~~\sigma^{\rm CC}(e^+p)~\sim~(1-y)^2xd_{\rm v}.
\ee 

\begin{figure}[t]
\begin{center}
\epsfxsize=15pc 
\epsfbox{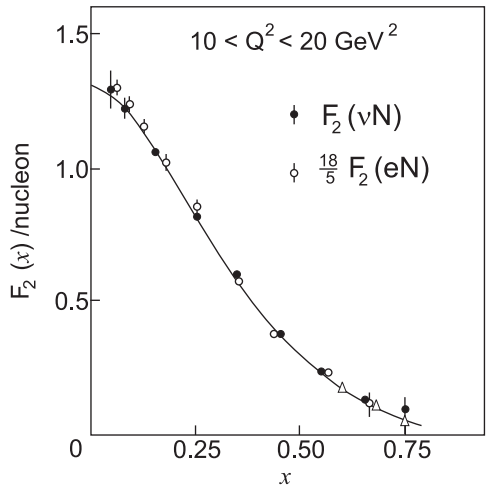} 
\caption{An historical plot of SLAC NC- and fixed-target muon neutrino CC-DIS data showing agreement with the QPM relations (\ref{eq:qpm3}) and (\ref{eq:qpm2}), but also indicating that only $50\%$ of the proton's momentum is carried by quarks.  \label{fig:comp}}
\end{center}
\end{figure}

It is informative to compare $F_2(N) \equiv (F_2^p+F_2^n)/2$ for NC and CC DIS. For NC we have from (\ref{eq:p}) and (\ref{eq:n})
\be
F_2^\gamma (eN) =\frac{5}{18} x(u+\bar{u}+d+\bar{d}+...), 
\label{eq:qpm3}
\ee
whereas for CC processes, $\nu N \to \mu X$, it follows from (\ref{eq:CCW}) that
\be
F_2^W (\nu N) = x(u+\bar{u}+d+\bar{d}+...). 
\label{eq:qpm2}
\ee
An experimental comparison of DIS data in the early 1970s is shown in Fig.~\ref{fig:comp}. The good agreement with the QPM relations is evident, but the area under the curve
\be
\int_0^1 F_2(\nu N)dx~=~\int_0^1 \sum_{q,\bar{q}} xq(x) dx~ \simeq ~0.5,
\ee
shows that only $50\%$ of the proton's momentum is carried by quarks. It provided the first (indirect) evidence for the existence of the gluonic component of the proton.

Before the advent of QCD there was a {\it big puzzle}. In DIS the struck quark appears to act as if it were free inside the proton. Yet it is never seen. No matter how hard it is hit, a free quark never emerges. It is confined within the proton. To say that a quark acts as if it were totally free, in a deep inelastic scatter, is not quite correct.  We need to allow for the interactions of quarks and gluons (QCD), and to see how this improves the QPM description of DIS. At the same time, we will see that QCD appears to be able to solve the {\it big puzzle}. It offers the possibility of an explanation of quark confinement.

\section{R\'{e}sum\'{e} of QCD}
Colour was first introduced to overcome the statistics problem in the quark model of hadron spectroscopy. Take for example the hadron $\Delta^{++}$, which was discovered as a resonance in $\pi^+p$ scattering. It has spin 3/2 and is made of three $u$ quarks, which can be in a state with all their spins parallel. But quarks are fermions and, by the exclusion principle, it should not be possible to have three identical quarks in the same state. A way to overcome the anomaly was soon proposed, which only later was found to have profound implications. The idea is to give quarks an additional attribute, {\it colour}, which can take three possible values, say red, green and blue. Hadrons are postulated to be {\it colourless} or, to be precise, colour singlets of the group SU(3) constructed from the fundamental colour triplet of quarks $(q_R,~q_G,~q_B)$. Baryons $(qqq)$ and mesons $(\qq)$ are clearly allowed, but single free quarks are forbidden, since they carry colour. In effect the puzzle of quark confinement (the experimental absence of free quarks) has been replaced by the puzzle of why colour should be confined.

\begin{figure}[t]
\begin{center}
\epsfxsize=18pc 
\epsfbox{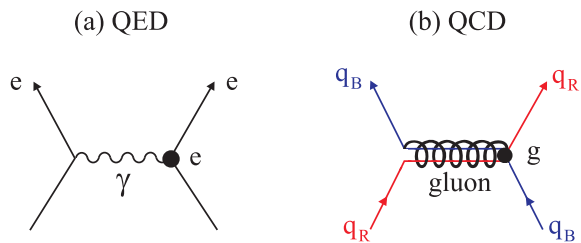} 
\caption{The lowest-order Feynman graphs for (a) the QED interaction between charged electrons and (b) the QCD interaction between coloured quarks via a coloured gluon, $(R\overline{B})$. \label{fig:qedqcd}}
\end{center}
\end{figure}

Recall that the electromagnetic force between charged particles is mediated by the exchange of photons, Fig.~\ref{fig:qedqcd}(a). The strength of the quantum electrodynamic (QED) interaction is determined by the charge of the particles. QED can be obtained from a remarkably simple symmetry principle: invariance of the theory under local phase transformations of the fields for the charged particles. Local means that we can arbitrarily vary the phase from space-time point to point. Since phase transformations commute with each other and form a U(1) symmetry group, we say that QED is a U(1) Abelian local gauge theory. The particles which emerge naturally from the theory to compensate for the phase differences from point-to-point or, in other words, to ensure the local gauge (i.e. phase) invariance of the theory, have zero mass and spin 1. These ``carriers'' of the electromagnetic force, the so-called gauge bosons, have exactly the properties of, and can be identified with, the familiar photon.

In analogy, in 1972, a local SU(3) gauge theory, quantum chromodynamics (QCD), was proposed as the theory of the strong interaction. In QCD the interaction is mediated by the exchange of zero-mass, spin-1 gluons between coloured quarks, Fig.~\ref{fig:qedqcd}(b). In QCD there are three different colour charges (red, green and blue) which have to be conserved, so the most general phase transformation is slightly more complicated. To be precise, QCD is based on invariance under the non-Abelian SU(3) group of local phase transformations among the triplet of colour charges, $q= (q_R,~q_G,~q_B)$.  The gluons themselves carry colour. In fact, eight different colour combinations of gluon are required to neutralize all possible phase differences: one colour combination is made explicit in Fig.~\ref{fig:qedqcd}(b).

Note that the gluons have zero-mass and therefore infinite range, and yet the strong force between hadrons has such a very short range. Indeed, it is ironic that the nuclear force, where it all began, is now relegated to a residual colour interaction, between colour neutral hadrons. The binding of colourless protons and neutrons into nuclei is similar to the van der Waals force which binds electrically neutral atoms into a molecule. Since colour is confined, the nuclear force must be short range and confined to hadronic dimensions.

\section{The running QCD coupling}
The most crucial feature of QCD is the dependence of the QCD coupling, $\alpha_s \equiv g^2/4\pi$, on $Q^2$. At first sight, it appears that a dimensionless\footnote{The structure function $F_2$ is such a dimensionless observable.} QCD observable $R$ must, for energies $Q \gg m_q$, be independent of $Q^2$.  The only energy scales in the QCD Lagrangian are the quark masses, and since the relevant ones are very light we would expect this scaling property to set in at low $Q^2$. However this argument is not true in a renormalizable field theory like QCD (or QED). A scale enters when we use perturbation theory to calculate the observable
\be
R~=~\sum_n c_n\alpha_s^n,
\ee
since we encounter (loop) Feynman diagrams which diverge logarithmically. We need to renormalize (reparameterize) the theory, which introduces a renormalisation scale $\mu$. As a consequence we find that the dimensionless observable $R$  no longer scales, but has logarithmic scaling violations, that is, it has the functional dependence $R({\rm log}(Q^2/\mu^2),\alpha_s (\mu^2))$.

\begin{figure}[t]
\begin{center}
\epsfxsize=28pc 
\epsfbox{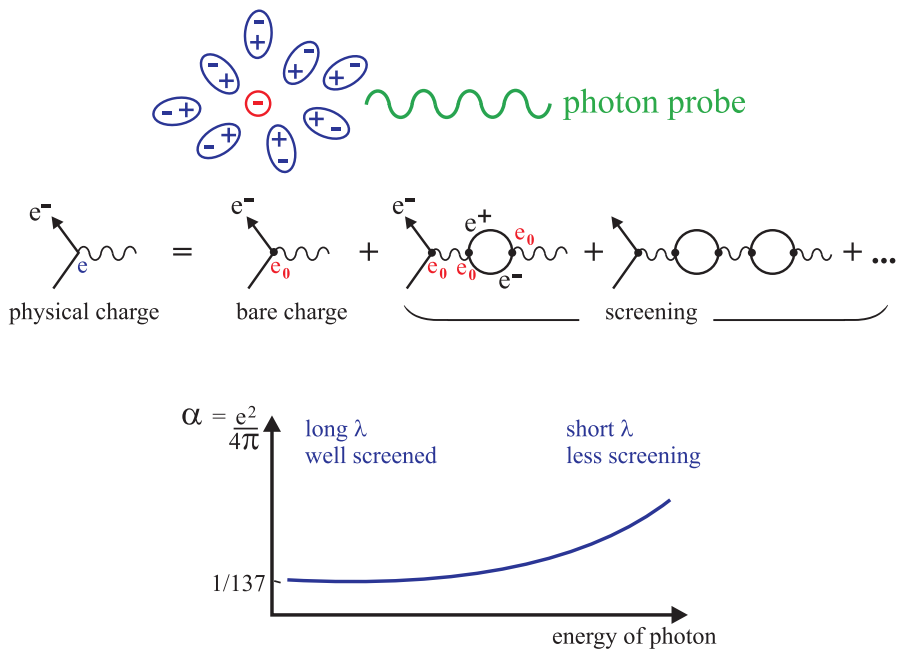} 
\caption{The vacuum polarisation effects which cause the QED coupling, $\alpha\equiv e^2/4\pi$, to run. The shorter the wavelength of the probing photon the more of the bare electron charge it sees. QED determines the running of the coupling $\alpha$, but experiment fixes the normalisation, which is traditionally given in terms of $\alpha(0)\simeq 1/137$. \label{fig:qedcoupl}}
\end{center}
\end{figure}

First we discuss the QED coupling. Vacuum polarisation effects (i.e. polarised $e^+e^-$-pairs) screen the bare electron charge.  The screening is least at short photon wavelengths, which causes the QED coupling, $\alpha=e^2/4\pi$, to increase with the energy of the photon. The situation is shown in Fig.~\ref{fig:qedcoupl}, which also shows the relevant Feynman diagrams. Summing up these diagrams we obtain, at large $Q^2$,
\be
\alpha(Q^2)=\alpha_0\left[1+\frac{\alpha_0}{3\pi}{\rm log}\frac{Q^2}{M^2}+\left(\frac{\alpha_0}{3\pi}{\rm log}\frac{Q^2}{M^2}\right)^2+...\right]=\frac{\alpha_0}{1-\frac{\alpha_0}{3\pi}{\rm log}\frac{Q^2}{M^2}}~,
\label{eq:loop}
\ee
where a cut-off, $M$, on the loop momentum has been introduced to prevent an infinite contribution. We may eliminate the dependence of $\alpha$ on this arbitrary parameter $M$ by introducing the renormalisation scale $\mu$. From (\ref{eq:loop}) we have
\be
\frac{1}{\alpha(Q^2)}=\frac{1}{\alpha_0}-\frac{1}{3\pi}{\rm log}\frac{Q^2}{M^2},~~~~~{\rm and}~~~~~\frac{1}{\alpha(\mu^2)}=\frac{1}{\alpha_0}-\frac{1}{3\pi}{\rm log}\frac{\mu^2}{M^2}.
\label{eq:QEDsub}
\ee
The $M$ dependence can be eliminated by subtracting these two relations. In this way we obtain
\be
\alpha(Q^2)~=~\frac{\alpha(\mu^2)}{1-(\alpha(\mu^2)/3\pi)~{\rm log}(Q^2/\mu^2)}~.
\label{eq:QEDrun}
\ee
In effect, the infinities of the theory have been removed at the price of introducing a renormalisation scale $\mu$. QED predicts the ``running'' of $\alpha$, but experiment is needed to predict its absolute value.

\begin{figure}[t]
\begin{center}
\epsfxsize=15pc 
\epsfbox{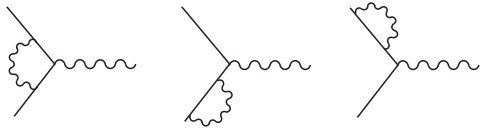} 
\caption{The Ward identities ensure that the ultraviolet divergences of these diagrams mutually cancel. \label{fig:ward}}
\end{center}
\end{figure}

Note that, due to basic properties of gauge field theories, the ultraviolet divergences of the Feynman diagrams of Fig.~\ref{fig:ward} mutually cancel, via the so-called Ward identities. This is just as well, because it ensures that the renormalized charge of the electron, muon,... remain equal.

\begin{figure}[t]
\begin{center}
\epsfxsize=28pc 
\epsfbox{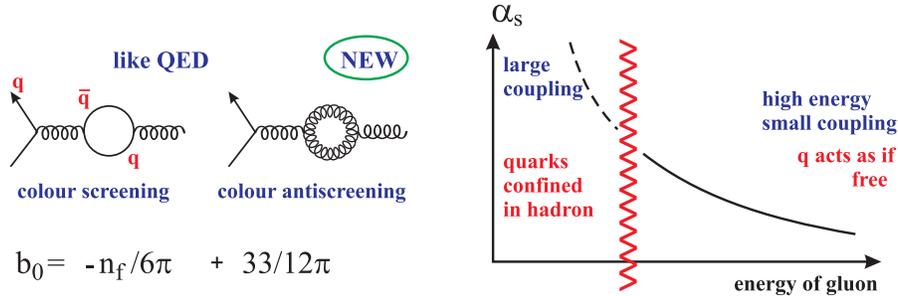} 
\caption{In addition to the `screening' quark loop, QCD has `antiscreening' from the gluon-loop diagram, which arises from the non-Abelian nature of the SU(3)-colour gauge group which, in turn, allows a triple-gluon vertex (as well as a quartic-gluon coupling, see Fig.~\ref{fig:2loops}). As a result the running of the QCD coupling constant, $\alpha_s$, is the `opposite' of QED. It decreases with energy, allowing the use of perturbation theory at high energies. \label{fig:QCDrun}}
\end{center}
\end{figure}

Turning now to QCD we have a new vertex to consider, the triple-gluon vertex, which arises since the gluons themselves carry colour charge. This changes everything, as Fig.~\ref{fig:QCDrun} shows. There is a new vacuum polarisation diagram with a gluon loop, which antiscreens the colour charge, which dominates the screening arising from the $n_f$ quark-loop diagrams\footnote{$n_f$ is the number of active quark flavours, that is the number of quarks with $m_q<Q$.}. As a consequence the $-1/3\pi$ in (\ref{eq:QEDrun}) becomes $+b_0$, with 
\be
\alpha_s(Q^2)~=~\frac{\alpha_s(\mu^2)}{1+b_0~\alpha_s(\mu^2)~{\rm log}(Q^2/\mu^2)}~.
\label{eq:QCDrunn}
\ee
where\footnote{The $-1/3\pi$ in (\ref{eq:QEDrun}) becomes $-1/6\pi$ for each quark loop in (\ref{eq:QCDrunn}) due to the convention used to normalize the SU(3) matrices.}
\be
b_0=-\frac{n_f}{6\pi}+\frac{33}{12\pi}.
\ee
At a stroke, the non-Abelian nature of QCD has solved the puzzling dilemma of the quark model. The big puzzle was that when the proton is hit hard in DIS, the quarks act as if they are essentially free; and yet no free quark has ever been seen -- they are confined within the hadron\footnote{Confinement still has to be proven. Lattice QCD is the technique to describe physics in the strong coupling regime.}. This {\it asymptotic freedom} and {\it infrared slavery} is precisely what the running of the QCD coupling indicates, see the $\alpha_s$ plot in Fig.~\ref{fig:QCDrun}.

The ultraviolet divergences of the QCD diagrams analogous to those in Fig.~\ref{fig:ward} cancel due to the Slavnov-Taylor identities of QCD. Moreover for a gauge theory the equality of the $\qq g$ and $ggg$ couplings is preserved by renormalization.

\begin{figure}[t]
\begin{center}
\epsfxsize=25pc 
\epsfbox{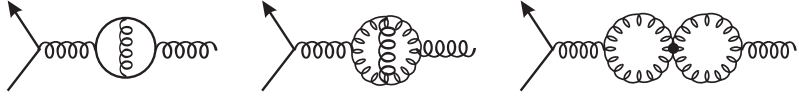} 
\caption{The diagrams which specify the two-loop $\beta$-function. Note that besides the triple-gluon vertex, QCD contains a quartic gluon coupling which gives rise to the final diagram. \label{fig:2loops}}
\end{center}
\end{figure}

Let us return to our dimensionless observable $R$, which due to renormalization becomes the function $R(Q^2/\mu^2,~\alpha_s(\mu^2))$. However $R$ cannot depend on the choice of renormalization scale, so we have a renormalization group equation (RGE)
\be
\frac{{\rm d} R}{{\rm dlog}\mu^2}~=\left(\frac{\partial}{\partial {\rm log}\mu^2}+\frac{\partial \alpha_s}{\partial {\rm log}\mu^2}\frac{\partial}{\partial\alpha_s}\right)R~=~0.
\label{eq:RGE}
\ee
It can be shown that the solution of the RGE gives
\be
R(Q^2/\mu^2,~\alpha_s(\mu^2))~=~R(1,~\alpha_s(Q^2)).
\label{eq:RGEsol}
\ee
That is the running of $\alpha_s$ determines the $Q$ dependence of $R$. In general, the running is expressed in terms of a $\beta$-function, defined by
\be
\partial \alpha_s/\partial{\rm log}~\mu^2~=~\beta (\alpha_s).
\label{eq:beta}
\ee
So far we have introduced the $\beta$-function at one-loop, in which we have summed up the leading logs, that is all the $(\alpha_s{\rm log}(Q^2/\mu^2))^n$ contributions.  From (\ref{eq:QCDrunn}) it is easy to show that this gives
\be
\beta(\alpha_s)~=~-b_0~\alpha_s^2.
\ee
The two-loop $\beta$-function sums up the next-to-leading logs resulting from the two-loop diagrams shown in Fig.~\ref{fig:2loops}, which are found to give an $\alpha_s^3$ term with coefficient $b_1=(153-19n_f)/24\pi^2$. The $\beta$-function then becomes
\be
\beta(\alpha_s)~=~-b_0~\alpha_s^2-b_1~\alpha_s^3.
\ee

Before returning to DIS and the structure of the proton, let us make a few more notes about the properties of the QCD coupling $\alpha_s$. First the coupling at two-loops is the solution of the transcendental equation
\be
\frac{1}{\alpha_s(Q^2)}-\frac{1}{\alpha_s(\mu^2)}+\frac{b_1}{b_0}{\rm log}\left(\frac{\alpha_s(Q^2)}{\alpha_s(\mu^2)}\left[\frac{b_0+b_1\alpha_s(\mu^2)}{b_0+b_1\alpha_s(Q^2)}\right]\right)~=~b_0{\rm log}\frac{Q^2}{\mu^2}.
\ee
Next, the value of $\alpha_s$ depends on the renormalization scheme. Since in two different schemes the values are related by $\alpha'_s=\alpha_s(1+c\alpha_s)$, it follows that $b_0$ and $b_1$ are scheme independent, whereas the higher coefficients are not. Nowadays, most calculations in fixed-order QCD perturbation theory are performed in the so-called modified minimal subtraction ($\overline{\rm MS}$) scheme. Thirdly, note that the coefficients $b_i$ depend on the number of active flavours, $n_f$. As $Q$ increases through a flavour threshold we will need to ensure the continuity\footnote{Beyond two-loops there are discontinuities in $\alpha_s$ at the flavour thresholds. Of course, the observables are continuous, since the above discontinuities are cancelled by ones occurring in the coefficients functions, see (\ref{eq:facth}) below.} of $\alpha_s$. For example at the $b$-quark threshold we will require $\alpha_s (m_b^2,4)=\alpha_s (m_b^2,5)$. Finally, recall that perturbative QCD tells us how the coupling varies with scale, but not the absolute value itself. The latter is obtained from experiment. Traditionally the value is quoted at $Q=M_Z$ in the $\overline{\rm MS}$ scheme for $n_f=5$; the current value, determined from many independent experiments is
\be
\alpha_s(M_Z^2)~=~0.117_6\pm 0.002.
\ee
Lattice QCD can also predict the value of the coupling. It is encouraging that these very different determinations are found to be consistent with each other.

\begin{figure}[t]
\begin{center}
\epsfxsize=27pc 
\epsfbox{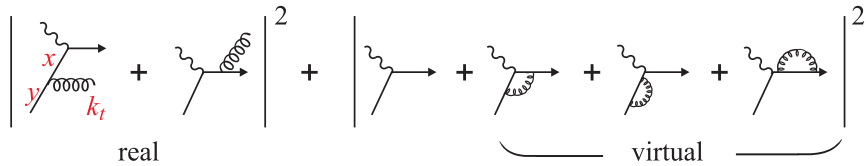} 
\caption{The $O(\alpha_s)$ diagrams which contribute to the proton structure function. On the first diagram we show new variables, the longitudinal momentum fraction $y$ carried by the quark and the transverse momentum $k_t$ of the emitted gluon, which must be integrated over. \label{fig:disqcd}}
\end{center}
\end{figure}
\section{Running parton densities: DGLAP equations}
In terms of QCD perturbation theory, the QPM formula (\ref{eq:QPMf}) may be regarded as the zeroth-order term in the expansion of $F_2$ as a power series in $\alpha_s$. To include the $O(\alpha_s)$ QCD corrections, we have to calculate the photon-parton subprocess diagrams shown in Fig.~\ref{fig:disqcd}. The QPM of Section 3 is the first diagram in the second $|...|^2$. Due to the propagator of the virtual quark, the first diagram in the first $|...|^2$ is proportional to $(yp-k)^{-2} \propto (2p \cdot k)^{-1}$, for massless quarks.  It therefore has a {\it collinear} divergence when the gluon of 4-momentum $k$ is emitted parallel to the incoming quark of 4-momentum $yp$. 

The partonic approach is only valid if the second diagram in the first $|...|^2$ can be neglected. Then, the emitted gluon can be considered as part of the proton structure. It turns out that both diagrams are required to ensure gauge invariance, but that the second only plays the role of cancelling the contributions from the unphysical polarization states of the gluon. Adopting a physical gauge, in which we sum only over transverse gluons, only the first diagram remains.

To evaluate its contribution we need to sum over all the possible values of the new variables, $y$ and $k_t$, that are introduced to describe the gluon. First we write down the answer and then explain its structure\footnote{This will be a prolonged discussion. Only in Section 8 will we emphasize the important ``factorization theorem'' structure of the formula.}. The result is
\be
\frac{F_2(x,Q^2)}{x}=\sum_q\int^1_x \frac{dy}{y} f_q(y) e^2_q\left[\delta\left(1-\frac{x}{y}\right)+\frac{\alpha_s}{2\pi} \left( P\left(\frac{x}{y}\right){\rm log}\frac{Q^2}{\mu_0^2}+C\left(\frac{x}{y}\right)  \right) \right],
\label{eq:f2qcd}
\ee
where $P$ and $C$ are known functions. These will become the universal parton splitting functions (here $P \equiv P_{qq}$ describes the $q \to qg$ splitting) and the process dependent coefficient function $C$. The $\delta$-function term in (\ref{eq:f2qcd}) is the zeroth-order QPM contribution with $y=x$ that we derived in Section 3. The first order term in $\alpha_s$ comes from the first diagram in Fig.~\ref{fig:disqcd}. A straightforward application of the Feynman rules shows that
\be
P(z)=\frac{4}{3}\frac{1+z^2}{1-z},
\label{eq:Pqq}
\ee
and that the log$(Q^2/\mu_0^2)$ originates from the integration over the gluon (bremsstrahlung) transverse momentum spectrum
\be
\int^{Q^2}_{\mu_0^2}\frac{dk^2_t}{k^2_t}~=~{\rm log}\left(\frac{Q^2}{\mu_0^2}\right),
\ee
where the upper limit is set by the virtuality of the photon which scatters off a quark of transverse size $1/Q$. Really the lower limit of integration should be set to zero. We have therefore arbitrarily cut-off the integral at some scale $\mu_0$. How do we make sense of (\ref{eq:f2qcd})?

Inspection of (\ref{eq:f2qcd}) shows that, after including the $O(\alpha_s)$ contribution, we may replace $f_q$ by a {\it well-behaved}\footnote{This follows because everything else in (\ref{eq:f2qcda}) is well-behaved.} running parton density 
\be
f_q(x,\mu^2)~=~f_q(x)~+~\int^1_x\frac{dy}{y}f_q(y)~\frac{\alpha_s}{2\pi}\left(P\left(\frac{x}{y}\right){\rm log}\left(\frac{\mu^2}{\mu_0^2}\right)+C_1\right),
\label{eq:frun}
\ee
such that
\be
\frac{F_2(x,Q^2)}{x}=\sum_q\int^1_x \frac{dy}{y} f_q(y,\mu^2) e^2_q\left[\delta\left(1-\frac{x}{y}\right)+\frac{\alpha_s}{2\pi} \left( P\left(\frac{x}{y}\right){\rm log}\frac{Q^2}{\mu^2}+C_2 \right) \right],
\label{eq:f2qcda}
\ee
where the division of the known function $C$ into $C_1+C_2$ depends on the choice of (factorization) scheme.

The dependence of $f_q(x,\mu^2)$ on the (non-perturbative) scale $\mu_0$ can be eliminated in an analogous way to the dependence of the coupling $\alpha$ on the cut-off $M$, see (\ref{eq:QEDsub}) and (\ref{eq:beta}). From (\ref{eq:frun}) we obtain
\be
\frac{\partial f_q(x,\mu^2)}{\partial {\rm log} \mu^2}~=~\frac{\alpha_s}{2\pi}\int^1_x\frac{dy}{y}f_q(y,\mu^2)~P\left(\frac{x}{y}\right),
\label{eq:evol}
\ee
which describes the evolution of the parton density with $\mu^2$. This is known as the DGLAP evolution equation \cite{dglap}. In effect we have absorbed all the collinear infrared sensitivity into a well defined {\it running} parton density $f_q(x,\mu^2)$. We cannot use perturbative QCD to calculate the absolute value of $f_q(x,\mu^2)$, but we can, via the DGLAP equation, determine its $\mu$ dependence.

\section{Further discussion of the DGLAP evolution equations}
Our $O(\alpha_s)$ treatment is incomplete. A hint that this is so, is the presence of a {\it soft} divergence, which arises when the energy of the emitted gluon tends to zero, in addition to the {\it collinear} divergence. This is reflected in the $z=1$ singularity of $P(z) \equiv P_{qq}(z)$, see (\ref{eq:Pqq}). At this point it is crucial to include the virtual gluon diagrams of Fig.~\ref{fig:disqcd}. The ($O(\alpha_s)$) contribution is the interference of these three diagrams with the QPM diagram. This contribution is also singular at $z=1$. It turns out that the singularity exactly cancels the $z=1$ singularity present in the real contribution. After the cancellation of the singularity there remains a residual $\delta (1-z)$ contribution from the virtual diagrams. Instead of calculating this contribution explicitly, there is an easy way to see what it must give. It must be such to satisfy 
\be
\int_0^1P_{qq}(z)~dz~=~0,
\label{eq:cons}
\ee
which expresses the fact that the number of valence quarks is conserved during the evolution. The virtual diagrams regularize the $1/(1-z)$ singularity in $P_{qq}$ so the constraint holds. This modification to $P_{qq}$ can be expressed in terms of the so-called ``+ prescription'' for regularization in which $1/(1-z)$ is replaced by $1/(1-z)_+$ defined so that
\be
\int_0^1 dz\frac{f(z)}{(1-z)_+}~=~\int_0^1 dz\frac{f(z)-f(1)}{(1-z)}
\ee
where $(1-z)_+=(1-z)$ for $z<1$. Ensuring the constraint (\ref{eq:cons}) gives
\be
P_{qq}~=~ \frac{4}{3}~\frac{1+z^2}{(1-z)_+}~+~2\delta(1-z).
\label{eq:pqqreg}
\ee

Our $O(\alpha_s)$ treatment is still not complete. In addition to the $\gamma q \to gq$ subprocesses shown in Fig.~\ref{fig:disqcd}, at $O(\alpha_s)$, we need to include the $\gamma g \to \qq$ processes. Then the DGLAP evolution equation (\ref{eq:evol}) for the quark density $q \equiv f_q$ becomes
\be
\frac{\partial q(x,Q^2)}{\partial {\rm log}Q^2}~=~\frac{\alpha_s}{2\pi}~\left( P_{qq} \otimes q~+~P_{qg} \otimes g \right)
\label{eq:glap1}
\ee
where $g \equiv f_g$ is the gluon density, and $P_{qq} \equiv P$ is the $q \to q(g)$ splitting function of (\ref{eq:pqqreg}). It can be shown that the $g \to q$ splitting function
\be
P_{qg}~=~\frac{1}{2}(z^2+(1-z)^2).
\ee
 In general $P_{ab}$ describes the $b \to a$ parton splitting. Also in (\ref{eq:glap1}) we have used $\otimes$ to abbreviate the convolution integral
\be
P \otimes f \equiv \int^1_x\frac{dy}{y}f_q(y)~P\left(\frac{x}{y}\right).
\ee
Clearly we must also consider the evolution of the gluon density
\be
\frac{\partial g(x,Q^2)}{\partial {\rm log}Q^2}~=~\frac{\alpha_s}{2\pi}~\left( \sum_i P_{gq} \otimes (q_i+\bar{q}_i)~+~P_{gg} \otimes g \right),
\ee
where the sum is over the $i$ quark flavours, and where the $q \to g$ and $g \to g$ splitting functions can be shown to be
\be
P_{gq}~=~P_{qq}(1-z)~=~\frac{4}{3}~\frac{1+(1-z)^2}{z},
\ee
\be
P_{gg}~=~6\left(\frac{1-z}{z}+\frac{z}{(1-z)_+}+z(1-z)\right)+\left(\frac{11}{2}-\frac{n_f}{3}\right) \delta (1-z).
\ee
Here the coefficient of the $\delta(1-z)$ term can be obtained from the constraint that all of the momentum of the proton must be carried by its constituents
\be
\int_0^1 dz~z\left(\sum_i (q_i(z,Q^2) + \bar{q}_i(z,Q^2))+g(z,Q^2)\right)~=~1
\label{eq:mom}
\ee
for all $Q^2$.

It is convenient to introduce flavour singlet $(\Sigma)$ and non-singlet $(q^{\rm NS})$ quark distributions:
\be
\Sigma~=~\sum_i (q_i+\bar{q}_i).
\ee
An example of a non-singlet is the up valence distribution
\be
u_{\rm v}=u-\bar{u}.
\ee
 Non-singlet evolution satisfies (\ref{eq:evol}) and decouples from the singlet and gluon evolution equations, which are coupled together as follows
\be
\frac{\partial}{\partial{\rm log}Q^2} \left( \begin{array}{c} \Sigma \\ g \end{array} \right) ~=~ \frac{\alpha_s}{2\pi}
\left( \begin{array}{cc} 
           P_{qq}~ & 2n_f P_{qg} \\
           P_{gq}~ & P_{gg}
           \end{array}    \right)
\otimes  \left( \begin{array}{c} \Sigma \\ g \end{array} \right). 
\ee

In general the splitting functions can be expressed as a power series in $\alpha_s$
\be
P_{ab}(\alpha_s,z)~=~P_{ab}^{\rm LO}(z)+\alpha_s P_{ab}^{\rm NLO}(z)+\alpha_s^2P_{ab}^{\rm NNLO}(z)+...
\ee     
where the NLO expressions were computed in the period 1977-80 and the NNLO in the period ending 2004. Leading order (LO) DGLAP evolution, which we have outlined, sums up the leading log contributions $(\alpha_s {\rm log}Q^2)^n$, and next-to-leading order evolution includes the summation of the $\alpha_s(\alpha_s {\rm log}Q^2)^{n-1}$ terms.

If we are given the $x$ dependence of the parton densities at some input scale $Q_0^2$ then we may solve the evolution equations to determine them at higher $Q^2$. Frequently this is performed simply by step-by-step integration up in $Q^2$. 

An alternative procedure is to rewrite the equations in terms of moments, which for an arbitrary function $f(z)$ are defined as
\be
f^{(n)}~=~\int_0^1 \frac{dz}{z}z^nf(z).
\ee
If we now multiply the DGLAP equation (\ref{eq:evol}) by $x^{n-1}$, and integrate over $x$, we obtain
\be
\frac{\partial }{\partial{\rm log}Q^2}\int^1_0 x^{n-1}q_{\rm NS}(x,Q^2)dx~=~\frac{\alpha_s}{2\pi}\int^1_0 y^{n-1}q_{\rm NS}(y,Q^2)dy~\int^1_0 z^{n-1} P_{qq}(z)dz
\ee
using $x=yz$. That is, the evolution equation then turns into an ordinary linear differential equation for the moments,
\be
\frac{\partial q^{(n)}_{\rm NS}}{\partial{\rm log}Q^2}~=~\frac{\alpha_s}{2\pi}P_{qq}^{(n)}q^{(n)}_{\rm NS}.
\label{eq:momt}
\ee
For fixed $\alpha_s$ the solution is
\be
q^{(n)}_{\rm NS}(Q^2)~=~c_n~{\rm exp}\left(\gamma^{(n)}{\rm log}Q^2\right)~=~c_n~[Q^2]^{\gamma^{(n)}},
\ee
where $\gamma^{(n)} \equiv \alpha_s P_{qq}^{(n)}/2\pi$ is known as the `anomalous dimension'. If we incorporate the running of $\alpha_s$, (\ref{eq:QCDrunn}), then it is easy to show that
\be
q^{(n)}_{\rm NS}(Q^2)~=~c_n~[\alpha_s(Q^2)]^{-\gamma^{(n)}/2\pi b_0}.
\ee
This is the LO behaviour. In analogy with (\ref{eq:RGE}) and (\ref{eq:RGEsol}), the general result may be obtained from the RGE 
\be
\frac{{\rm d} q^{(n)}}{{\rm dlog}\mu^2}~=\left(\frac{\partial}{\partial {\rm log}\mu^2}+\beta(\alpha_s)\frac{\partial}{\partial\alpha_s}+\gamma^{(n)}(\alpha_s)\right)q^{(n)}~=~0,
\label{eq:RGEmom}
\ee
which can be shown to have the solution
\be
q^{(n)}(Q^2/\mu^2,\alpha_s(\mu^2))~=~q^{(n)}(1,\alpha_s(Q^2))~{\rm exp}\left(\int^{\alpha_s(Q^2)}_{\alpha_s(\mu^2)}\frac{\gamma^{(n)}(\alpha_s)}{\beta(\alpha_s)}d\alpha_s\right).
\ee

In addition to (\ref{eq:momt}) we have
\be
\frac{\partial}{\partial{\rm log}Q^2} \left( \begin{array}{c} \Sigma^{(n)} \\ g^{(n)} \end{array} \right) ~=~ \frac{\alpha_s}{2\pi}
\left( \begin{array}{cc} 
           P_{qq}^{(n)}~ & 2n_f P_{qg}^{(n)} \\
           P_{gq}^{(n)}~ & P_{gg}^{(n)}
           \end{array}    \right)
  \left( \begin{array}{c} \Sigma^{(n)} \\ g^{(n)} \end{array} \right). 
\ee
Once we have the analytic solutions of these equations for the moments, we can obtain the $z$ distributions of the partons by the inverse Mellin transforms
\be
f_i(z,Q^2)~=~\frac{1}{2\pi i}\int_{c-i\infty}^{c+i\infty} dn~z^{-n}f_i(n,Q^2),
\ee
where the contour is to the right of all the singularities of the integrand.

\section{Observables: the factorization theorem}
We return to the equation for $F_2(x,Q^2)$, eq.(\ref{eq:f2qcda}). We have described how the collinear singularities of the formula have been swept into well-defined {\it running} parton densities, $f_i(y,\mu^2_F)$, evaluated at some (factorization) scale\footnote{The subscript $F$ is added to distinguish it from the renormalization scale introduced in Section 5. In practice these scales are often chosen to be equal.} $\mu_F$ in the perturbative region.  A convenient choice is to set $\mu_F=Q$, so that the log$(Q^2/\mu^2_F)$ term disappears. We then have, including the $\gamma g \to \qq$ contribution,
\begin{eqnarray}
\frac{F_2(x,Q^2)}{x} &  = &  \sum_{q,\bar{q}}e^2_q\int^1_0 \frac{dy}{y} \left[ f_q(y,Q^2) \left(\delta\left(1-\frac{x}{y}\right)+ \frac{\alpha_s}{2\pi}C_{2,q}\left(\frac{x}{y}\right)  \right) \right.  \nonumber \\
 &  &  \qquad \qquad \qquad + \left. f_g(y,Q^2)\frac{\alpha_s}{2\pi}C_{2,g}\left(\frac{x}{y}\right) \right],
\end{eqnarray}
where the $C_{2,i}$ are the coefficient functions for the observable $F_2$. Although all the collinear singularities are absorbed by the running of the $f_i$, recall that the prescription is not unique. We can add any finite term. So we must specify a scheme. The $\overline{\rm MS}$ factorization scheme is favoured. It was mentioned at the end of Section 5 as also the choice of renormalization scheme.

\begin{figure}[t]
\begin{center}
\epsfxsize=27pc 
\epsfbox{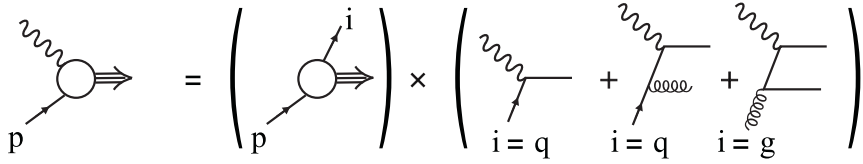} 
\caption{Schematic picture of the factorization theorem for a deep inelastic structure function of the proton. \label{fig:fac}}
\end{center}
\end{figure}

We can generalize this result to describe the structure functions of all DIS processes. For the structure functions $F_a$, describing the deep inelastic processes $\ell+p \to \ell'+X$, the factorization formula, which holds to all orders in perturbation theory, has the generic form
\be
F_a(x,Q^2)~=~\sum_{i=q,\bar{q},g}\int_0^1\frac{dy}{y}~f_i(y,Q^2)~C_{a,i}\left(\frac{x}{y},\alpha_s(Q^2)\right)~~+~~O\left(\frac{\Lambda^2_{\rm QCD}}{Q^2}\right).
\label{eq:facth}
\ee
The final term denotes non-perturbative contributions, such as hadronization effects, multiparton interactions etc. For sufficiently high $Q^2$ these effects are negligible, and the expression for the observable factorizes into
\begin{itemize}
\item {\it universal} parton densities (of the proton), $f_i$, which absorb the {\it long distance} collinear singularities. They cannot be calculated in perturbative QCD, but their $Q^2$ dependence is calculable using the DGLAP evolution equations, in which the splitting functions are calculable as power series in $\alpha_s$.
\item coefficient functions, $C_{a,i}$, which describe the {\it short distance} subprocess. They are calculable from perturbative QCD as a power series in $\alpha_s$, but are unique to the particular observable, $F_a$.
\end{itemize}
The factorization is displayed visually in Fig.~\ref{fig:fac}

A similar factorization applies to inclusive `hard' hadron-hadron collisions. For instance, consider the LHC process
\be
p(p_1)~+~p(p_2)~ \to ~H(Q,..)~+~X
\ee
where $H$ denotes the triggered hard system, such as a weak boson, a pair of jets, a Higgs boson etc. The typical hard scale $Q$ could be the invariant mass of $H$ or the transverse momentum of a jet. Then according to the factorization theorem the cross section is of the form
\be
\sigma~=~\sum_{i,j}\int^1_{x_{\rm min}}dx_1dx_2 ~f_i(x_1,\mu_F^2) f_j(x_2,\mu_F^2) ~\hat{\sigma}_{ij}(x_1p_1,x_2p_2,Q...;\mu_F^2),
\ee
where typically $x_{\rm min} \gapproxeq Q^2/s$ where $s=(p_1+p_2)^2$. For the production of a system $H$ of invariant mass $M$ and rapidity $y$, the momentum fractions $x_{1,2}=Me^{\pm y}/\sqrt{s}$. The $f_i$ and $\hat{\sigma}$ depend on the renormalization scale $\mu_R$ via $\alpha_s(\mu_R^2)$. For instance
\be
\hat{\sigma}_{ij}~=~\alpha_s^k\sum_{m=0}^n C^{(m)}_{ij}\alpha_s^m 
\ee
where LO, NLO... correspond to $n$=1,2...; note that, for example, $k$=0,2,.. for $W$, dijet,... production. We should work to the same order in the series expansion of the splitting functions. In practical applications it is usual to choose $\mu_F=\mu_R \sim Q$ and to use variations about this value to estimate the uncertainty in the predictions.
Of course the physical cross section $\sigma$ does not depend on the scales, but the truncation of the perturbative series brings in scale dependence. If we truncate at order $\alpha_s^n$, then the uncertainty is of order $\alpha_s^{n+1}$.

\section{Global parton analyses}
Two groups (CTEQ \cite{cteq} and MRST \cite{mstw}) have used all available deep inelastic and related hard scattering data involving incoming protons (and antiprotons) to determine the parton densities, $f_i$, of the proton.  The procedure is to parametrize the $x$ dependence of $f_i(x,Q^2_0)$ at some low, yet perturbative, scale $Q^2_0$. Then to use the DGLAP equations to evolve the $f_i$ up in $Q^2$, and to fit to all the available data (DIS structure functions, Drell-Yan production, Tevatron jet and $W$ production...) to determine the values of the input parameters.  In principle there are 11 parton distributions $(u, \bar{u}, d, \bar{d}, s, \bar{s}, c, \bar{c}, b, \bar{b}, g)$. However $m_c, m_b \gg \Lambda_{\rm QCD}$. So $c=\bar{c}$ and $b=\bar{b}$ are calculated from perturbative QCD via $g \to Q\bar{Q}$. Also the evidence from neutrino-produced dimuon data, $\nu N \to \mu^+\mu^-X$, is that\footnote{Analysis of NuTeV data for $\nu$ and $\bar{\nu}$ beams indicates some $x$ dependence of the factor ``0.2'', and that $s > \bar{s}$ for $x \sim 0.01$. } $s \simeq \bar{s} \simeq 0.2(\bar{u}+\bar{d})$ at $ Q^2 \simeq 1~{\rm GeV}^2$.

A common choice of parametrization of the parton densities is
\be
xf(x,Q^2_0)~=~A(1-x)^\beta x^\alpha (1+\epsilon\sqrt{x}+\gamma x)
\ee
with up to five parameters $(A,\alpha,\beta,\epsilon,\gamma)$ for each parton. Three of the $A$'s are determined from sum rules. The input partons must satisfy the two valence quark sum rules
\be
\int_0^1 dx(u-\bar{u})=2, ~~~~~~~~~\int_0^1 dx(d-\bar{d})=1,
\ee
and also we must satisfy the momentum sum rule (\ref{eq:mom}). 

We can obtain some idea of what to expect for the values of the $\beta_i$ parameters from the spectator counting rules. As $x \to 1$ physical arguments indicate that 
\be
f(x)~\to~(1-x)^{2n_s-1}
\ee
where $n_s$ is the minimum number of spectator quarks which share between them the residual, vanishingly small momentum of the proton. The greater the number of spectators, the smaller the chance of producing a parton with a large fraction of the proton's momentum. For a valence quark, gluon and sea quark it is easy to see that we have $n_s=2,3 $ and 4 respectively. So we may expect $\beta_{\rm v} \sim3, ~\beta_g \sim 5$ and $\beta_{\rm sea} \sim 7$. 

For a rough guide to the anticipated values of the $\alpha_i$ parameters, we might appeal to Regge behaviour, since the limit $x=Q^2/2p\cdot q \to 0$ corresponds to $s_{\gamma p} \simeq 2p\cdot q \to \infty$. In this limit the $\gamma p$ cross section is approximately proportional to
\be
\sum e_i^2 xf_i(x) \sim (r_P s_{\gamma p}^{\alpha_P(0)-1}+r_R s_{\gamma p}^{\alpha_R(0)-1}) \sim (r_P x^{1-\alpha_P(0)}+r_R x^{1-\alpha_R(0)}).
\ee
The naive expectations are that the Pomeron and the leading secondary Reggeons have trajectories with intercepts $\alpha_P(0) \simeq 1.08$ and $\alpha_R(0) \simeq 0.5$. The Pomeron corresponds to flavourless exchange so we expect the parameters $\alpha_{{\rm sea},g} \sim -0.08$, whereas the valence density corresponds to flavour exchange with $\alpha_{\rm v} \sim 0.5$. So, in summary, we might naively expect 
\be
xf_{\rm v} \sim x^{0.5} (1-x)^3,~~~~xf_g \sim x^{-0.08} (1-x)^5,~~~~xf_{\rm sea} \sim x^{-0.08} (1-x)^7
\label{eq:input}
\ee
types of behaviour.

In practice, the heavy quark densities, $c,b$, require special treatment. These are particularly important at small $x$, especially as $Q^2$ increases. We can see the problem by noting that for $Q^2 \sim m_c^2$ the charm quark does not act like a parton, but instead is created in the final state by photon-gluon fusion, $\gamma g \to \cc$. On the other hand for  $Q^2 \gg m_c^2$, clearly $c$ behaves like a massless parton. It is therefore necessary to use a variable flavour number scheme \cite{rst} in which we match a 3- to a 4-flavour parton description as we evolve up through the charm quark threshold, $Q^2 \sim m_c^2$.

\begin{figure}
\begin{center}
\epsfxsize=32pc 
\epsfbox{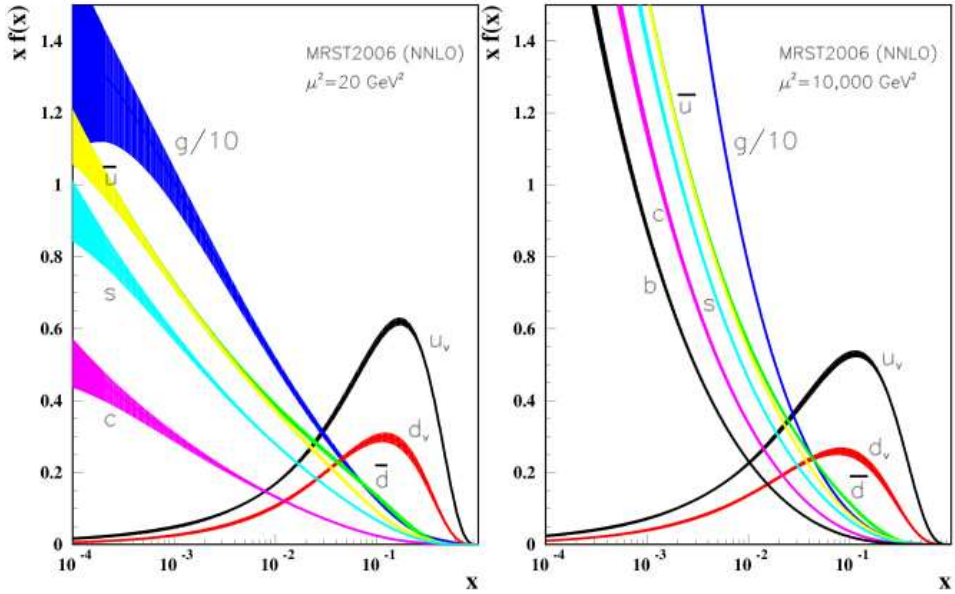} 
\caption{Parton densities, $xf_i(x,\mu^2)$, at $\mu^2=20$ and $10^4~\GeV^2$, obtained in a recent NNLO global analysis \cite{mstw}. The dominance of the gluon at small $x$ and of the valence quarks at large $x$ is clearly evident. The uncertainties shown only reflect the errors of the experimental data. A discussion of the theoretical errors can be found in \cite{mrste2}.} 
\label{fig:mrst}
\end{center}
\end{figure}
\begin{figure}
\begin{center}
\epsfxsize=15pc 
\epsfbox{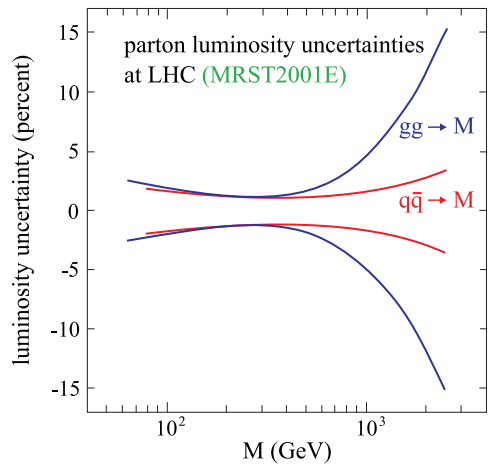} 
\caption{The uncertainty in the $\qq$ and $gg$ parton luminosities for producing a state of mass $M$ at the LHC, arising from the experimental errors of the data fitted in a global parton analysis.}
\label{fig:ggqqX}
\end{center}
\end{figure}
\begin{table}
\caption{ Lepton-nucleon and related hard-scattering processes (whose data are used in the global parton analyses) and their primary sensitivity
to the parton distributions that are probed.}
\begin{center}
\begin{tabular}{lcc}
\hline
\hline
& Main & PDFs \\
~~~~Process & Subprocess & probed \\ \hline

${  \ell^\pm N \rightarrow \ell^\pm X }$&  $\gamma^*q\rightarrow q$ & $g(x \lapproxeq 0.01),q,\overline{q}$ \\
${
\ell^+ (\ell^-) N \rightarrow \overline{\nu} (\nu) X }$ & $W^* q\rightarrow q'$  & " \\
${ \nu (\overline{\nu}) N
\rightarrow \ell^- (\ell^+) X}$ & $W^* q\rightarrow q'$ & "\\
${ \nu \;N \rightarrow \mu^+\mu^-X}$ &
$W^*s\rightarrow c\rightarrow \mu^+$  & $s$ \\
${  \ell N \rightarrow \ell QX  }$ & $\gamma^*Q\rightarrow Q$ &  $Q=c,b$ \\
         &  $\gamma^* g\rightarrow Q\overline{Q} $ &  $g(x \lapproxeq 0.01)$ \\
${ pp \rightarrow \gamma X}$ & $qg\rightarrow \gamma q$ & $g$ \\
 $pN \to \mu^+\mu^-X$ & $q\overline{q} \to \gamma^*$ & $\overline{q}$ \\
 ${ pp,pn \rightarrow \mu^+\mu^- X}$ & $u\overline{u},d\overline{d}\rightarrow \gamma^*$ &
$\overline{u}-\overline{d}$ \\
                               & $u\overline{d},d\overline{u}\rightarrow \gamma^*$ &  \\
${ ep,en \rightarrow e \pi X}$ & $\gamma^*q \rightarrow q$ & \\
 ${ p\overline{p} \rightarrow W \rightarrow
\ell^{\pm}X}$ &$ud\rightarrow W$ & $u,d,u/d$ \\
${ p\overline{p} \rightarrow}$ { jet} ${ +X}$ &
$gg,qg,qq\rightarrow 2j$&  $q,g (0.01 \lapproxeq x \lapproxeq 0.5)$ \\
\hline
\hline
\end{tabular}
\end{center}

\end{table}
Table 1 highlights some processes used in the global fits, and their primary sensitivity to the parton densities. The kinematic ranges of the fixed-target and collider experiments are complementary (as is shown in Fig.~\ref{fig:kin}) which enables the parton densities to be determined over a wide range in $x$ and $Q^2$. The analyses can now be done to NNLO. An example\footnote{Comprehensive sets of parton densities available as programme-callable functions can be found in http://durpdg.dur.ac.uk/HEPDATA/PDF.} of the resulting parton distributions is shown in Fig.~\ref{fig:mrst}. 

The gluon density is the most poorly known parton distribution. At small $x~(\lapproxeq 0.01)$ it is constrained by the HERA DIS scaling violations, and for values of $x$ up to about 0.5 by the Tevatron jet data. The momentum sum rule also gives an important constraint. 

\begin{figure} [t]
\begin{center} 
\epsfxsize=25pc 
\epsfbox{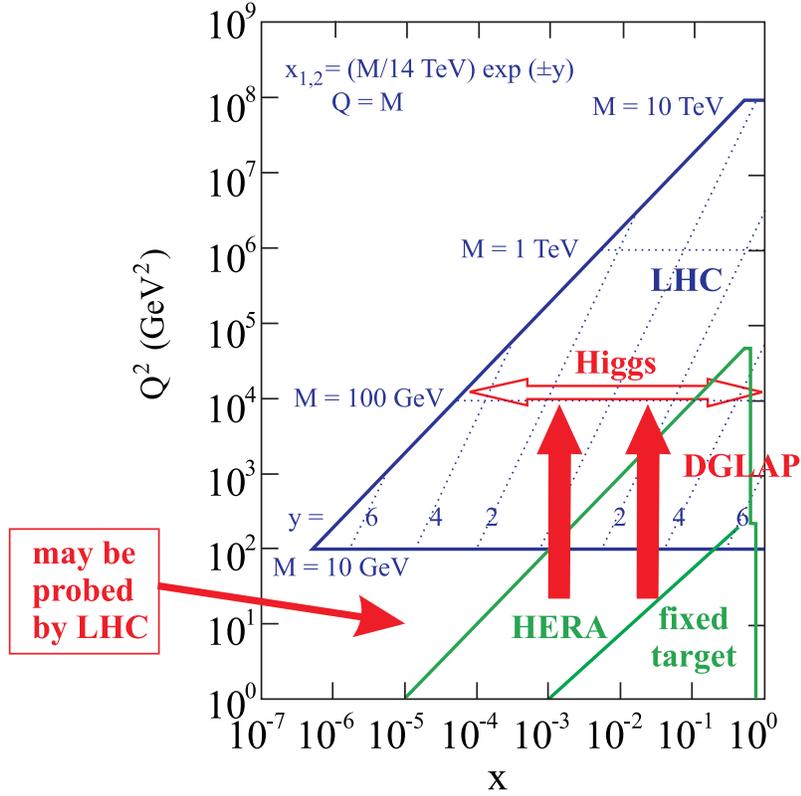} 
\caption{Partonic $x,Q^2$ domains sampled by the LHC and HERA, as well as fixed-target DIS experiments. The rapidity interval for the production of a Higgs boson of mass 120 GeV at the LHC is indicated by an open arrow; the relevant parton distributions should be reliable from DGLAP evolution of global analyses of HERA, fixed-target DIS, and Tevatron jet data. The possibility of the LHC experiments probing the region $x \lapproxeq 10^{-4},~Q^2 \gapproxeq 10~{\rm GeV}^2$ is mentioned at the end of Section 12.}
\label{fig:LHCkin}
\end{center}
\end{figure}
Thanks to the HERA experiments, the parton densities are well-known\footnote{We discuss possible corrections arising from the resummation of log$1/x$ terms and from absorptive effects, both of which lie outside pure DGLAP, in Sections 10 and 11 respectively. We shall see that, at low scales, the parton densities have large uncertainties for $x \lapproxeq 10^{-3}$.} down to about $x \sim 10^{-3}$. Also they are well-known up to $x \sim 0.5$.  What are the implications of the uncertainties\footnote{Detailed discussions of the uncertainties arising in the global analyses can be found in \cite{cteqe1,cteqe2,mrste1,mrste2}.} in the parton densities for the LHC experiments? Some idea can be obtained from Fig.~\ref{fig:ggqqX}, which shows the uncertainties in the ${\cal L}_{\qq}$ and ${\cal L}_{gg}$ parton luminosities relevant to the production of a state of mass $M$ at the LHC. The parton luminosities are defined as
\be 
{\cal L}_{ab}~=~C_{ab}\int^1_\tau \frac{dx_a}{x_a} f_a(x_a)f_b(\tau/x_a) 
\ee
where $C_{ab}$ is a colour factor. Since $x_a x_b s \simeq M^2$ we see $x_b=\tau/x_a$ where $\tau=M^2/s$. Due to the factorization theorem, the cross section for the production of the state of mass $M$ is
\be
\sigma~=~\sum_{a,b} {\cal L}_{ab} ~\hat{\sigma}(ab \to M;\hat{s}=\tau s).
\ee
The widening of the $gg \to M$ error band in Fig.~\ref{fig:ggqqX} for $M>1$ TeV is due to the lack of knowledge of the gluon at high $x$. This plot does not include the theoretical errors in a pure DGLAP parton analysis. Nevertheless, for the predictions of the cross sections of the central production of high mass systems at the LHC, the uncertainty coming from parton densities is less than $\pm 10\%$. This is also clear from an inspection of Fig.~\ref{fig:LHCkin}. 

\begin{figure}
\begin{center}
\epsfxsize=20pc 
\epsfbox{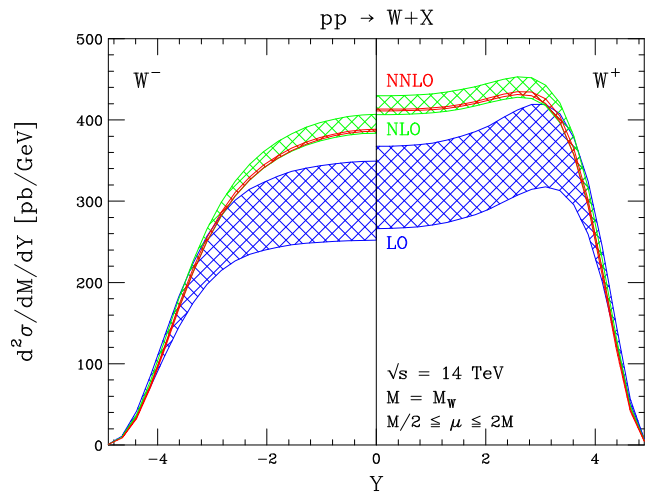} 
\caption{LO, NLO and NNLO predictions for the rapidity distribution of $W^{\pm}$ production at the LHC. The width of the bands reflects the uncertainty coming from the variation of the scale in the interval $M_W/2 \le \mu \le 2M_W$. The NNLO prediction is the very narrow band lying within the NLO error band.} 
\label{fig:W}
\end{center}
\end{figure}

As an example, we show in Fig.~\ref{fig:W} the predicted cross sections\cite{admp} for $W^{\pm}$ production at the LHC. At zeroth order we only have the $\qq$-driven subprocesses $u\bar{d} \to W^+$ and $d\bar{u} \to W^-$; so we expect the parton luminosity errors to be relatively small.  The cross section inequality $\sigma(W^+)>\sigma(W^-)$ reflects $u_{\rm v}>d_{\rm v}$. Also note the rapid decrease in the uncertainty due to scale changes as we proceed from LO$\to$NLO$\to$NNLO. Allowing for uncertainties from all sources, the $W^{\pm}$ production cross section is predicted to an accuracy of $\pm 5\%$, which enables it to be considered as a luminosity monitor for the LHC.

\begin{figure}
\begin{center}
\epsfig{file=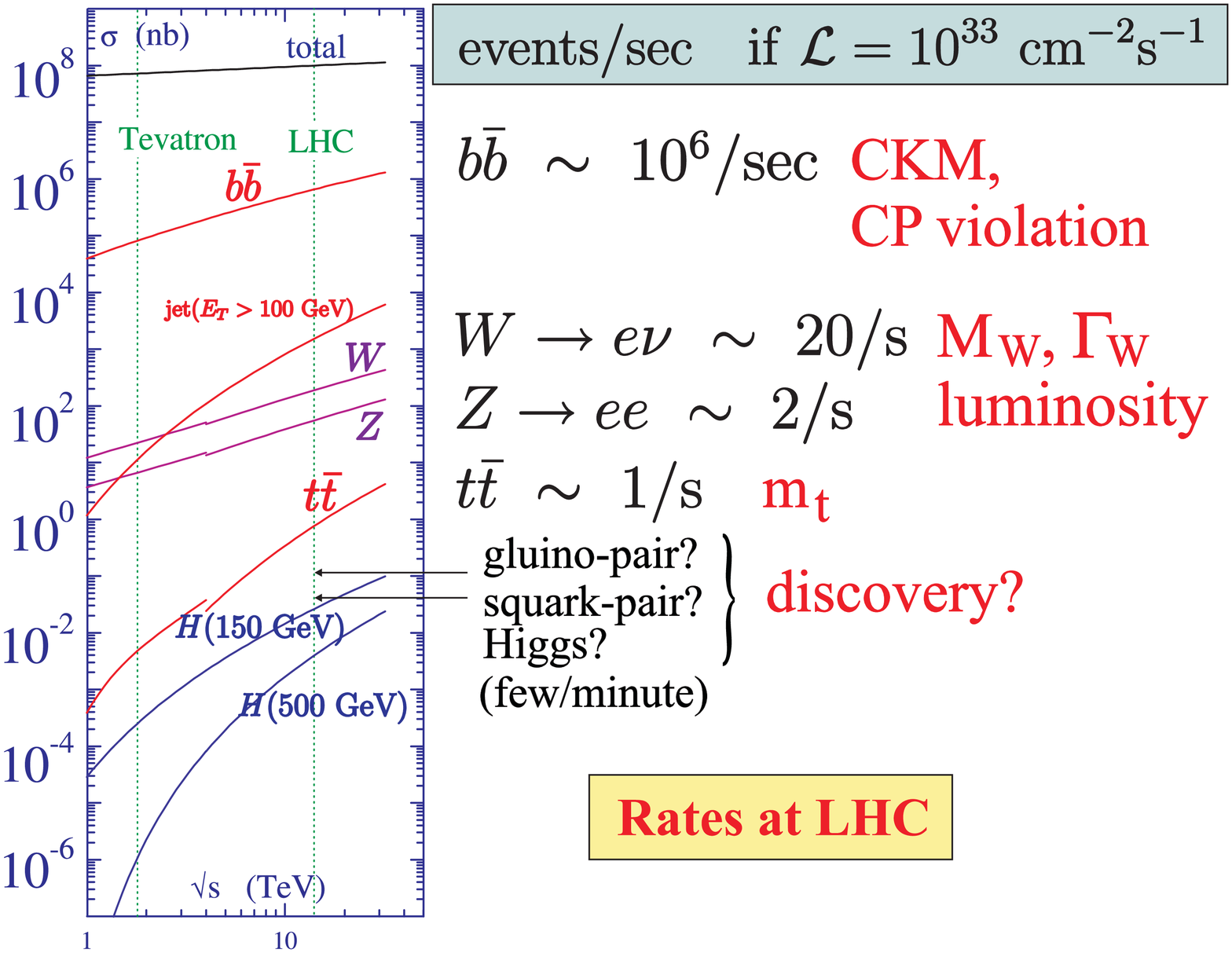,width=14cm,angle=0}
\end{center}
\caption{The cross sections (in nb) for various processes at the Tevatron and the LHC. For the LHC luminosity quoted, the scale also corresponds to the number of events/second. We also give an indication of the physics which may be probed by the processes at the LHC. Note that the rates of Higgs and SUSY particle production do not include the dilution of a possible signal due to the branching fraction of the particular channel investigated. Moreover note how important it is to reduce the huge background and to overcome ``pile-up'' from multiple events per bunch crossing at the higher luminosity. Of course it would be even more exciting to discover something totally unexpected.} 
\label{fig:LHCrates}
\end{figure}

In Fig.~\ref{fig:LHCrates} we show the cross sections in nb for various processes at the Tevatron and at the LHC. If the collider luminosities were $10^{33} ~{\rm cm}^{-2} {\rm s}^{-1}$, then the scale on the plot also gives the number of events which would occur each second. Note that eventually the LHC is planned to achieve a luminosity some 10 times greater than this.

\begin{figure} [t]
\begin{center}
\epsfig{file=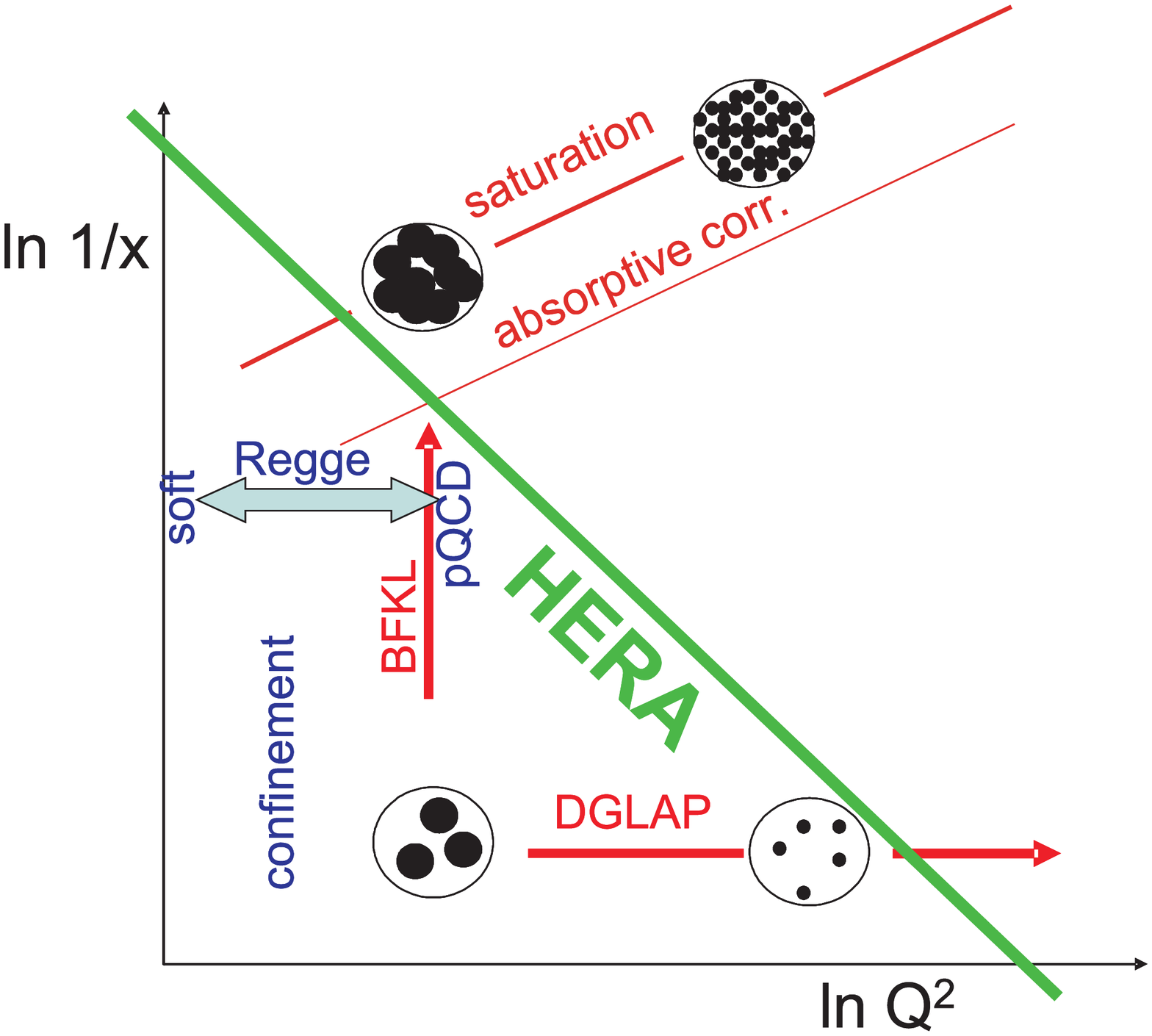,width=10cm,angle=0} 
\end{center}
\caption{Schematic sketch of the physical phenomena in various regions of the log$(1/x)$ -- log$(Q^2)$ plane, compared to the kinematic reach of HERA. The gluonic content of the proton, as resolved by a $Q^2$ probe, is also indicated. DGLAP evolution takes us up in $Q^2$ and so the partonic constituents are resolved more finely. The BFKL equation takes us to small $x$, with the gluon density $xg$ growing as $x^{-\lambda}$, but the resolution in the transverse plane remaining at approximately $1/Q$. As $x$ decreases, the partonic content increases, and at some stage the partons recombine (absorptive effects), and eventually saturate.} 
\label{fig:hera}
\end{figure}

\section{Beyond DGLAP: low $x$ partons and BFKL}
Fig.~\ref{fig:hera} shows the physical phenomena we expect to be appropriate in various regions of the log$(1/x)$ -- log$(Q^2)$ plane. We shall discuss them here and in the next Section. Overlaid is a line indicating the reach achieved by the HERA experiments. Of course the position of this line is well known, see Fig.~\ref{fig:kin}. However the positions of the various domains relative to this line are not well established. Certainly HERA has opened up the small $x$ domain, with DIS structure function measurements reaching down to $x \sim 10^{-4}$ while $Q^2$ is still in the perturbative domain. 

So far our approach has been to work with DGLAP evolution truncated at a fixed perturbative order. This pure DGLAP approach has been phenomenologically successful, even, surprisingly, down to $x \sim 10^{-4}$ with $Q^2 \sim 2~\GeV^2$. Nevertheless, although the global parton analyses describe the data satisfactorily in this regime\footnote{The gluon has a {\it valence-like} behaviour, although the uncertainties are large in this domain. Nevertheless its behaviour is quite different to the growth of the sea-quark distributions as $x \to 0$. Such a result looks strange from the Regge viewpoint where the same
vacuum singularity (Pomeron) should drive both the sea quarks and the gluons;
i.e. the same small $x$ behaviour is expected for sea quark
and gluon distributions.}, it does not mean that the parton distributions are reliable here. We know pure DGLAP is incomplete at small enough $x$. 

To explore the small $x$ regime, we first note that DGLAP is equivalent to assuming that the dominant dynamical mechanism leading to DIS scaling violations is the evolution of parton emissions strongly-ordered in transverse momenta. However, at small $x$ the evolution occurs over large rapidity intervals ($\sim {\rm ln}1/x$). The higher-order corrections to the splitting (and coefficient) functions contain one additional power of ln$1/x$ for each additional power of $\alpha_s$. If we keep just the leading ln$1/x$ terms then the small $x$ behaviour of the $P_{gg}$ splitting function, for example, has the form
\be 
xP_{gg}(x)~\to~ A_{10}~ \alpha_s + A_{21}~\alpha_s^2 {\rm ln}1/x + A_{32}~ \alpha^3_s {\rm ln}^2 1/x +A_{43}~\alpha^4_s {\rm ln}^3 1/x  +...~,
\label{eq:pggbfkl}
\ee
whereas in NNLO DGLAP, for example, $P_{gg}$ contains only the terms up to $\alpha_s^3$. Clearly, at small $x$, when $\alpha_s~{\rm ln} 1/x \sim 1$, a resummation of all of the terms in the series is necessary. The resummation of the leading log (${\rm LL}_x$) terms, $\alpha_s^n {\rm log}^{n-1}1/x$, is accomplished by the BFKL equation\footnote{ DGLAP and BFKL are different limits of a more general evolution of parton densities, which is an ordered evolution in the angles of the emitted partons. At LO we have strong ordering of the emission angles,...$\theta_i \ll \theta_{i+1}$...; on the other hand if, at one step of the evolution $\theta_i \sim \theta_{i+1}$, then this contribution is included inside the NLO splitting function. In the collinear approximation of DGLAP the angle increases due to the growth of the transverse momentum $k_t$, while in BFKL the angle ($\theta \simeq k_t/k_\parallel$) grows due to the decreasing longitudinal momentum fraction as we proceed along the emission chain from the proton. Introductory discussions of the BFKL equation can be found, for example, in Refs.\cite{dd,fr}.}. The BFKL equation \cite{bfkl} will be discussed in detail in the lectures of Victor Fadin \cite{fad}, Lev Lipatov \cite{lev} and Al Mueller \cite{mu}. Here we will just include some introductory remarks.

At low $x$ we have diffusion or ``random walk'' in the logarithm of transverse momenta as we proceed along the emission chain. We no longer have the strong ordering in $k_t$ which is true in DGLAP evolution.  For this reason the BFKL equation is for the gluon density, $f(x,k_t^2)$, unintegrated over $k_t$. Recall that the gluon dominates at low $x$. The BFKL equation has the structure
\be
\frac{\partial f_g}{\partial {\rm ln}(1/x)}~=~K \otimes f_g~=~\lambda f_g
\ee
which, at small $x$, has the the solution
\be
f_g~ \sim ~ e^{\lambda {\rm ln}(1/x)}~\sim ~x^{-\lambda}~\sim ~\left(\frac{s}{s_0}\right)^\lambda,
\label{eq:lamb}
\ee
where $\lambda=12\alpha_s{\rm ln}2/\pi$ is the leading eigenvalue of the BFKL kernel $K$. This has an analogous form to the Regge-pole exchange behaviour of the amplitude,
\be
A(s,t)~ \sim ~ \sum_R \beta_R(t)\left(\frac{s}{s_0}\right)^{\alpha_R(t)},
\label{eq:a1}
\ee
which is the cornerstone of the description of high-energy ``soft'' hadron-hadron interactions; $\alpha_R(t)$ is the trajectory of Reggeon $R$ in the complex angular momentum plane. For colour-octet exchange the BFKL equation describes a Reggeized gluon with trajectory $\alpha_g(t)$, while for colour-singlet exchange, which is relevant to this discussion, it leads to a cut in the complex angular momentum, $j$, plane corresponding to two Reggeized gluons exchanged -- often called the perturbative Pomeron. Note that the generalized gluon distribution $f_g$ corresponds to the two-gluon exchange amplitude. Its behaviour at low $x$ is driven by the rightmost singularity (branch point), $j=1+\lambda$, produced by the two-gluon cut, where the value of $\lambda$ obtained from the BFKL equation is given above. Since the behaviour of $f_g$ is driven by a cut (and not an isolated pole) in the $j$-plane, a prefactor $1/\sqrt{{\rm ln}s}$ will appear in (\ref{eq:lamb}).  The possible connection between (\ref{eq:a1}) and (\ref{eq:lamb}) is indicated by a horizontal `block' arrow in Fig.~\ref{fig:hera}. In the ``soft'' regime the hadrons are Reggeized, while in the perturbative QCD BFKL regime the constituent partons are Reggeized. How to go from one regime to the other has not been solved. For example, what is the relation\footnote{See \cite{pvl} for a phenomenological study.} of the BFKL or perturbative QCD `Pomeron' (given by a ladder diagram formed from the exchange of two $t$-channel Reggeized gluons) to the `Pomeron' describing soft high-energy proton-proton interactions?

Coming back to the discussion centred on the perturbative expansion of equation (\ref{eq:pggbfkl}), we note that in the small $x$ region the gluon dominates, and only $P_{gg}$ and $P_{qg}$ contain ${\rm LL}_x$ contributions. These are positive but smaller than naively expected; it turns out that $A_{21}=A_{32}=0$, and even $A_{54}=0$, in (\ref{eq:pggbfkl}). Now the next-to-leading (${\rm NLL}_x$) terms, $\alpha^n_s{\rm log}^{n-2}1/x$, have also been calculated \cite{fadlip}. These give a large, negative, contribution to the gluon, leading to instability at small $x$. In fact $\lambda$ of (\ref{eq:lamb}) is now given by
\be
\lambda=12\alpha_s{\rm ln}2/\pi~(1-6.5\alpha_s).
\ee
This problem has been the subject of considerable investigation. Clearly, the higher-order contributions, ${\rm NNLL}_x$, ${\rm NNNLL}_x$,... are important. However it took about 10 years to calculate the ${\rm NLL}_x$ contributions, so to compute the next order or two appears unrealistic, and even then may not converge to a stable result. Instead, the procedure that has been followed is to identify a few physical QCD effects that lead to large higher-order corrections and then to resum them.  Indeed, this all-order resummation of the main effects is found to tame the wild (${\rm LL}_x$ $\to$ ${\rm NLL}_x$) behaviour; a readable review is given in \cite{salam}. The approaches of the various groups have reached similar conclusions: the approximate all-order resummed BFKL framework leads to the behaviour that
\be
xg \sim x^{-0.3} ~~~~~ {\rm as} ~~~~x \to 0
\label{eq:sx}
\ee
at low scales\footnote{What do the data say? If the $F_2$ data are fitted to the form $x^{-\lambda}$ for $x<0.01$, then it is found that $\lambda$ grows approximately linearly with log$Q^2$ from $\lambda \simeq 0.1$ passing through $\lambda =0.3$ at $Q^2 \sim 40~\GeV^2$. The simple assumption that this reflects the behaviour of the gluon, with $F_2$ driven by the $g \to q\bar{q}$ transition is much too naive. Indeed the global analyses give a gluon which is valence-like at small $x$ at the input scale.}. 

In practice, it is found that this power-like growth only sets in at very small $x$.  In terms of DGLAP evolution all the BFKL effects should be included in the resummed splitting functions used to describe the transition between two quite different scales, that is between partons whose transverse momentum are very different. In such a case the power growth (\ref{eq:sx}) will be included in the resummed $P_{gg}$. However the resummed $xP_{gg}$ has a {\it dip} centred at $x \sim 10^{-3}$, and the power growth is only evident below $x \sim 10^{-5}$. Indeed the resummed $xP_{gg}$ and the NNLO DGLAP $xP_{gg}$ are in good agreement down to $x \sim 10^{-3}$.

 To make quantitative predictions in the small $x$ domain, $x \lapproxeq 10^{-4}$ with $Q^2 \sim 2~\GeV^2$, where no data exist, is extremely difficult. We need to obtain the resummed ln$(1/x)$ solution starting from some non-perturbative amplitude at $Q=Q_0$. This non-perturbative distribution (which is analogous to the `input' in the DGLAP approach) is not known theoretically. Either one has to fit it to data (but again low $x$ data are needed) or to use some phenomenological model (for example, based on a Regge parametrization).  

We conclude that the parton densities are unknown in the region $x \lapproxeq 10^{-4}$.  At very small $x$ we have the estimate that gluon density might behave as $xg \sim x^{-\lambda}$ with $\lambda \simeq 0.3$. However, as $x$ decreases, at some stage this behaviour will violate unitarity. Here the recombination of gluons (absorptive effects) come to the rescue, and tame the violations of unitarity. To this we now turn.

\section{Absorptive effects}
The saturation of parton densities ($\lambda=0$) may be obtained using the Gribov-Levin-Ryskin (GLR) equation \cite{GLR} or the more precise Balitski-Kovchegov (BK) equation \cite{BK}. These equations sum up the set of so-called {\it fan}
diagrams which describe the rescattering of intermediate partons on
the target nucleon. The screening caused by these rescatterings prohibits
the power growth of the parton densities. 

The GLR equation for the gluon may be written in the symbolic form
\be
\frac{\partial(xg)}{\partial{\rm ln}Q^2}~=~P_{gg} \otimes g+P_{gq} \otimes q -\frac{81\alpha_s^2}{16R^2Q^2} \int \frac{dx'}{x'}[x'g(x',Q^2)]^2.
\ee
The non-linear shadowing term, $-[g]^2$, describes the recombination of gluons. It arises from perturbative QCD diagrams which couple $4g$ to
$2g$ --- that is two gluon ladders recombining into a single gluon ladder, which is called a {\it fan} diagram.  
The minus sign occurs because the scattering amplitude corresponding to a gluon ladder is 
predominantly imaginary. 
The parameter $R$ is a measure of the transverse area $\pi R^2$ where the gluons are concentrated.   

The BK equation is an improved version of the GLR equation.
 It accounts for a more precise form of the triple-pomeron vertex and can be
used for the non-forward amplitude. The GLR equation, based on DGLAP evolution, was in momentum space; whereas the BK equation, based on the BFKL equation, is written in coordinate
space in terms of the dipole scattering amplitude $N(\xb,\yb,Y)\,\equiv\, N_{\xb\yb}(Y)$. Here $\xb$ and $\yb$ are the transverse coordinates of the two $t$-channel
gluons which form the colour-singlet dipole, and $Y={\rm ln}(1/x)$ is the rapidity. The BK equation reads
\be    
\label{BKe}    
\frac{\partial N_{\xb\yb}}{\partial Y}\,=\,    
\frac{3\alpha_s}{\pi}\!    
\int\frac{d^2\zb}{2\pi}    
\frac{(\xb-\yb)^2}{(\xb-\zb)^2(\yb-\zb)^2}\,    
\left\{    
N_{\xb\zb}+N_{\yb\zb}-N_{\xb\yb}    
- N_{\xb\zb}\,N_{\yb\zb}   
\right\}\, ,    
\ee    
where, interestingly, the non-linear and linear terms have the {\it same} BFKL kernel $K$, which is shown explicitly in (\ref{BKe}). For small dipole densities, $N$, the quadratic term in the brackets may be
neglected, and, indeed, (\ref{BKe}) reproduces the conventional BFKL equation.
However for large $N$, that is $N\to 1$, the r.h.s. of (\ref{BKe}) vanishes,
and we reach saturation when $N=1$.
The equation sums up the set of fan diagrams where at the lower
(small $Y$) end the target emits any number of pomerons
 (i.e. linear BFKL amplitudes), while at the upper (large $Y$) end
we have only one BFKL dipole. 

In principle, it would appear more appropriate to use the BFKL-based BK equation to describe the parton
densities at low $x$. It is an attempt to describe 
 saturation phenomena. However it is just a model and cannot, at present, be used to
 reliably estimate absorptive effects at small $x$. 

Is there any evidence of the onset of absorptive effects in the experimental data? These should occur first at low $x$ and low $Q^2$, see Fig.~\ref{fig:hera}. However, there is no conclusive evidence that absorptive effects are important in the HERA data in the perturbative regime, $Q^2 \gapproxeq 1~\GeV^2$. The various claims that are frequently made have been recently comprehensively discussed in \cite{watt}. It is seen that none of them, including the observed `flat' ratio of (diffractive DIS/inclusive DIS) or the observation of geometric scaling, provide any compelling evidence of saturation effects.

Of course, as $x$ decreases we know that ultimately absorptive effects must be present. In principle, we should be able to estimate their contribution from knowledge of the structure functions for diffractive DIS, via
\be
\Delta F_2^{\rm abs}~\sim~ -F_2^{\it D},
\ee
where $F_2^D$ is the structure function for the process $\gamma^* p \to X+p$ in which the slightly deflected proton and the cluster $X$ of outgoing hadrons are well separated in rapidity \cite{mrw}.

\section{Conclusions}

The great improvement in the precision and range of deep inelastic and related hard scattering data over the last few years has enabled the partonic structure of the proton to be well determined in the $10^{-3} \lapproxeq x \lapproxeq 0.5$ interval, so we are able to make reliable predictions for the production of new massive states at the LHC. Global analyses are now available at NNLO. These analyses require particularly careful treatment at the heavy quark thresholds, see Thorne \cite{rst} and references therein. A surprise is that a pure DGLAP description is able to describe all features of the data down to $Q^2=2~\GeV^2$, in spite of the fact that the global fits are now quite tightly constrained.  The allowance of beyond-DGLAP effects is not found to improve the description, see, for example \cite {mrste2}.

Another surprise is that the global analyses reveal that the gluon has a valence-like small $x$ behaviour at the low input scale, unlike the sea quark distribution which behaves as expected, see (\ref{eq:input}).
Such a result looks strange from the Regge viewpoint where the same
vacuum singularity (Pomeron) should drive both the sea quarks and the gluons;
i.e. the same power $\lambda_g=\lambda_{\rm sea}$ is expected for sea quarks
and gluons. Note that global analyses are only reliable down to $x \sim 10^{-3}$.

The low $x$ domain, $x \lapproxeq 10^{-4}$, is unchartered territory. Is it possible for the LHC experiments to determine the behaviour of partons in the important low $x$ region below $10^{-4}$ at low scales? 
One possibility is $\mu^+ \mu^-$ Drell-Yan production in which events are observed with the
$\mu^+ \mu^-$ invariant mass as low as possible and the rapidity as large as possible.  For example, for $M_{\mu \mu}=4$ GeV and $y_{\mu \mu}=3$, we sample quarks at $x=1.4 \times 10^{-5}$, see Fig.~\ref{fig:LHCkin}. This process samples predominantly the sea quark distributions.  To study the small
$x$ behaviour of the gluon at low scales we may consider $\chi_c$ production\footnote{In practice, rather than $\chi_c$, it may be better to observe the process $pp \to J/\psi ~X$ as a function of $y_{J/\psi}$. This process also depends sensitively on the gluon distribution via the subprocesses $gg \to J/\psi~g,~~gg \to \chi \to J/\psi~\gamma$. There are also contributions from the subprocesses $gg \to b\bar{b}$ with $b \to J/\psi$, and $q\bar{q} \to J/\psi$. The analysis of such data will be considerably helped by the detailed observations of prompt $J/\psi$, as well as $J/\psi$ from $b$, central production at the Tevatron.}, or prompt photon production driven by the subprocess  $gq \to \gamma q$, or perhaps $gg \to b\bar{b}$. These studies may also require an improvement in theoretical formalism.

\section*{Acknowledgements}
I thank Dick Roberts, Misha Ryskin, James Stirling, Robert Thorne and Graeme Watt for many fruitful discussions and enjoyable collaborations on the subject of partons. I also thank Roberto Fiore and Christophe Royon for organizing such an enjoyable School.

\end{document}